\documentclass[12pt]{article}
\usepackage{epsfig,graphicx,color}
\usepackage{amsmath}
\usepackage{latexsym}
\usepackage{amstext}
\usepackage{epsfig, graphicx}
\newcommand{\mb}[1]{ \mbox{\boldmath$#1$} }
\newcommand{\ds}{\displaystyle}
\newcommand{\beq}{\begin{eqnarray}}
\newcommand{\eeq}{\end{eqnarray}}
\newcommand{\beqq}{\begin{eqnarray*}}
\newcommand{\eeqq}{\end{eqnarray*}}
\newcommand{\p}{\partial}

\newcommand{\eps}{\varepsilon}
\newcommand{\x}{\mbox{\boldmath$x$}}
\newcommand{\n}{\mbox{\boldmath$n$}}

\newcommand{\w}{\mbox{\boldmath$w$}}
\newcommand{\y}{\mbox{\boldmath$y$}}

\newcommand{\A}{\mbox{\boldmath$A$}}

\newcommand{\AB}{\mbox{\boldmath$AB$}}
\font\bb=msbm10 at 12pt
\def\rR{\hbox{\bb R}}

\begin{document}
\pagestyle{plain}
\begin{center}
{\large \textbf{{Brownian Motion In Dire Straits}}}\\[5mm]
D. Holcman\footnote{Department of Applied Mathematics, Tel-Aviv
University, Tel-Aviv 69978, Israel. Department of Mathematics and
Computational Biophysics, Ecole Normale Sup\'erieure, 46 rue d'Ulm
75005 Paris, France. This research is supported by an
ERC-starting-Grant.}, Z. Schuss\footnote{Department of Mathematics,
Tel-Aviv University, Tel-Aviv 69978, Israel.}
\end{center}
\date{}
\begin{abstract}
The passage of Brownian motion through a bottleneck in a bounded domain is a
rare event and as the bottleneck radius shrinks to zero the mean time for such
passage increases indefinitely. Its calculation reveals the effect of geometry
and smoothness on the flux through the bottleneck. We find new behavior of the
narrow escape time through bottlenecks in planar and spatial domains and on a
surface. Some applications in cellular biology and neurobiology are discussed.
\end{abstract}

\section{Introduction}
\begin{figure}
\centering {\includegraphics[width=4cm]{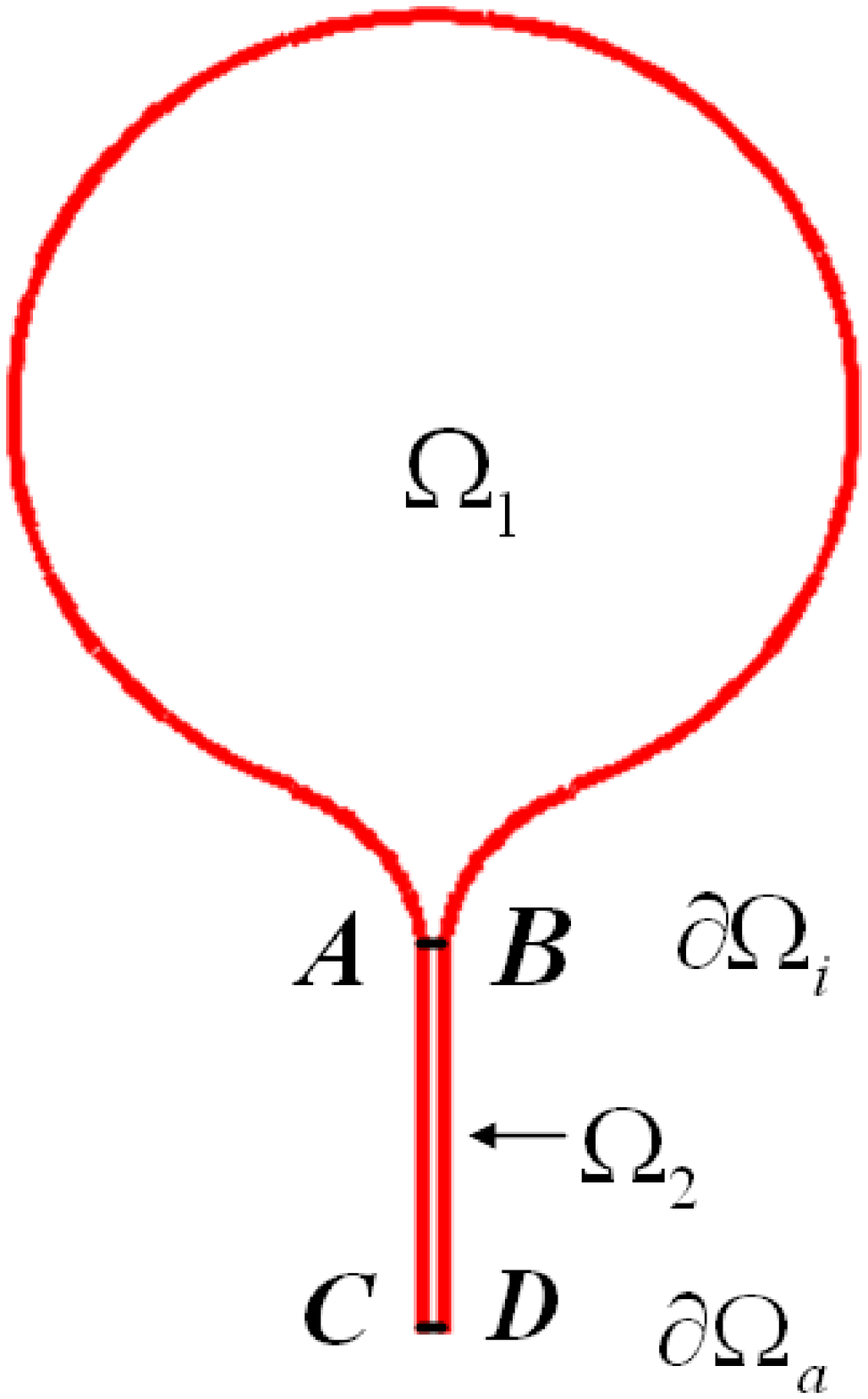}}
{\includegraphics[width=4.2cm]{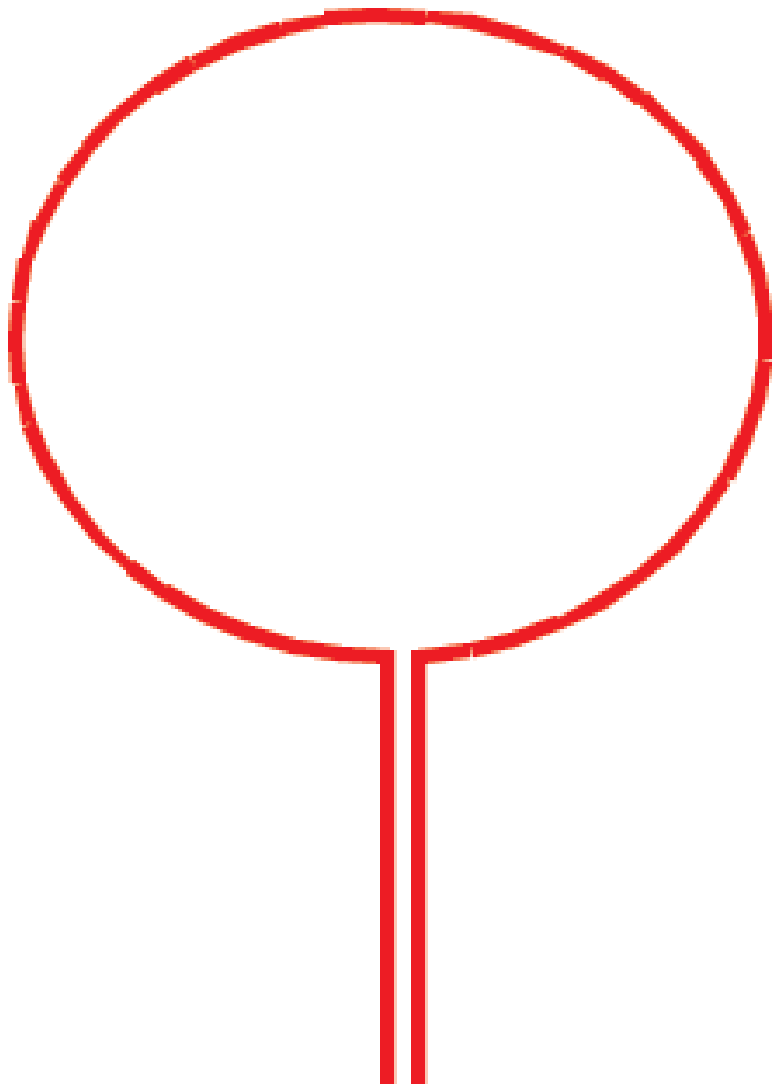}} \caption{\small A mathematical
idealization of a cross section of smooth and sharp connections approximating
the spine morphology: {\bf Left:} The cross section is a composite domain that
consists of a bulky head $\Omega_1$ connected smoothly by an interface
$\p\Omega_i=AB$ to a narrow neck $\Omega_2$. The entire boundary is
$\p\Omega_r$ (reflecting), except for a small absorbing part $\p\Omega_a=CD$.
{\bf Right:} A cross section of a sharp connection.} \label{f:sharp-spine}
\end{figure}

The narrow escape problem is to calculate the mean first passage time (MFPT) of
Brownian motion from a domain with mostly reflecting boundary to a small
absorbing window. The MFPT, also known as the narrow escape time (NET), was
calculated in \cite{Ward1}-\cite{PNAS} for small absorbing windows in a smooth
reflecting boundary. Several more complex cases were considered in
\cite{SSH1}-\cite{SSH3}, such as the NET through a window at a corner or at a
cusp in the boundary and the NET on Riemannian manifolds. The calculation of
the NET in composite domains with long necks, as shown in Figure
\ref{f:sharp-spine}, was attempted in \cite{PNAS} and \cite{Berez} and
ultimately accomplished in \cite{Dire-Part-II}. The NET problem in a planar
domain with an absorbing window at the end of a funnel was considered in
\cite{HHS-PRE}. The case of planar domains that consist of large compartments
interconnected by funnel-shaped bottlenecks was also considered in
\cite{HHS-PRE}.

\begin{figure}
\centering
\resizebox{!}{6cm}{\includegraphics{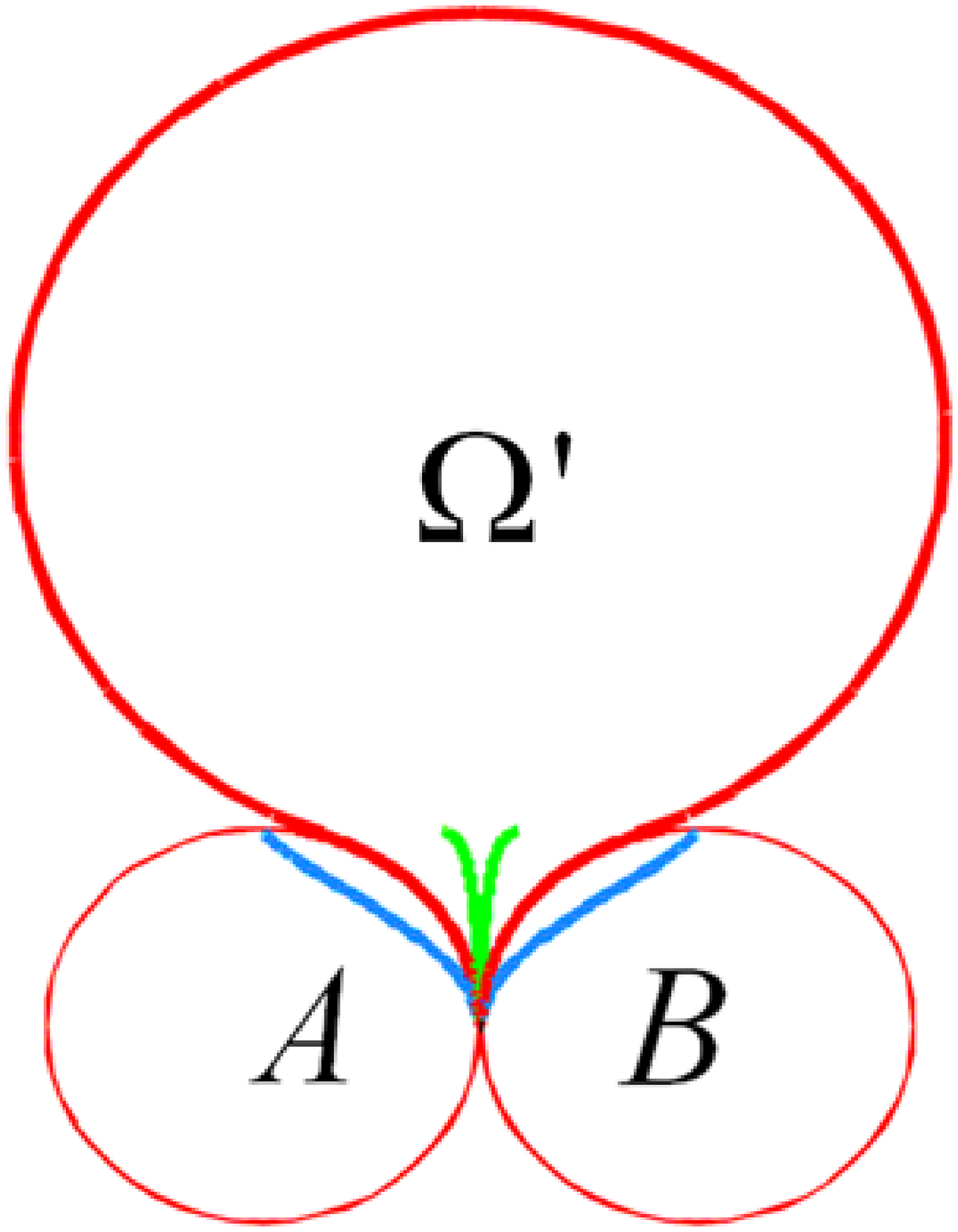}}\resizebox{!}{5.5cm}{\includegraphics{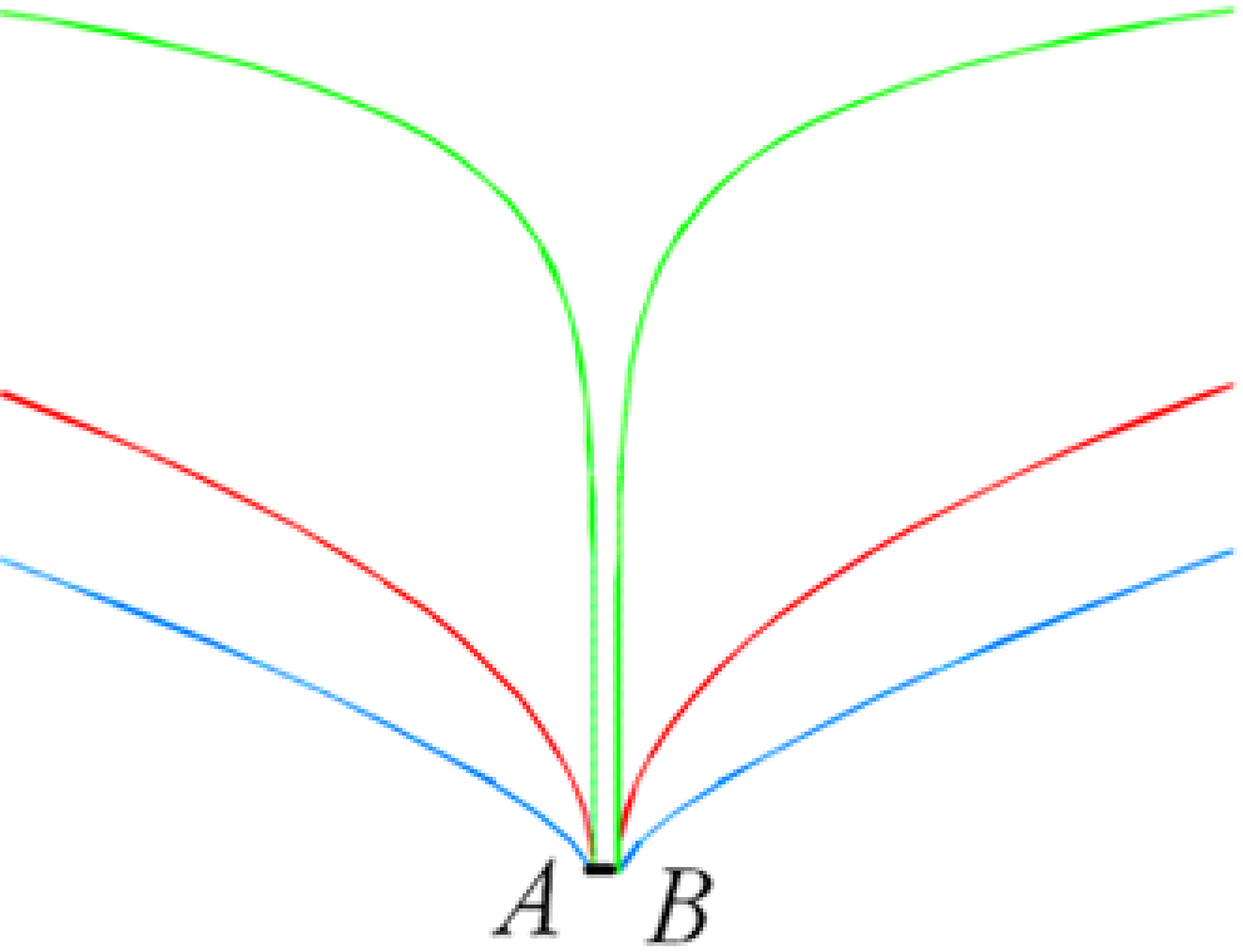}}
\caption{\small {\bf Left:} The planar (dimensional) domain $\Omega'$ is
bounded by a large circular arc connected smoothly to a funnel formed by moving
$\eps$ apart two tangent circular arcs of radius $R_c$ (i.e.,
$\overline{AB}=\eps$). {\bf Right:} Blowup of the cusp region. The red, green,
and blue necks correspond to $\nu_\pm=1,\,0.4$, and $5$ in (\ref{rzL}),
respectively.}\label{f:Canonical-partial-block1}
\end{figure}

\begin{figure}
\centering \resizebox{!}{5cm}{\includegraphics{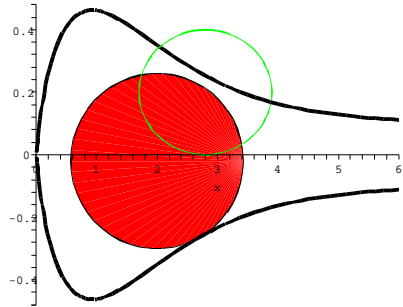}} \caption{\small Narrow straits
formed by a partial block (solid disk) of the passage from the head to the neck of the domain
enclosed by the black line. Inside the green circle the narrow straits can be approximated by the
gap between adjacent circles.} \label{f:Partial-block}
\end{figure}

In this paper we consider Brownian motion in two- and
three-dimensional domains whose boundaries are smooth and
reflecting, except for a small absorbing window at the end of a
cusp-shaped funnel, as shown in Figure
\ref{f:Canonical-partial-block1}. The cusp can be formed by a
partial block of a planar domain, as shown in Figure
\ref{f:Partial-block}.
\begin{figure}
onr\centering \resizebox{!}{8cm}{\includegraphics{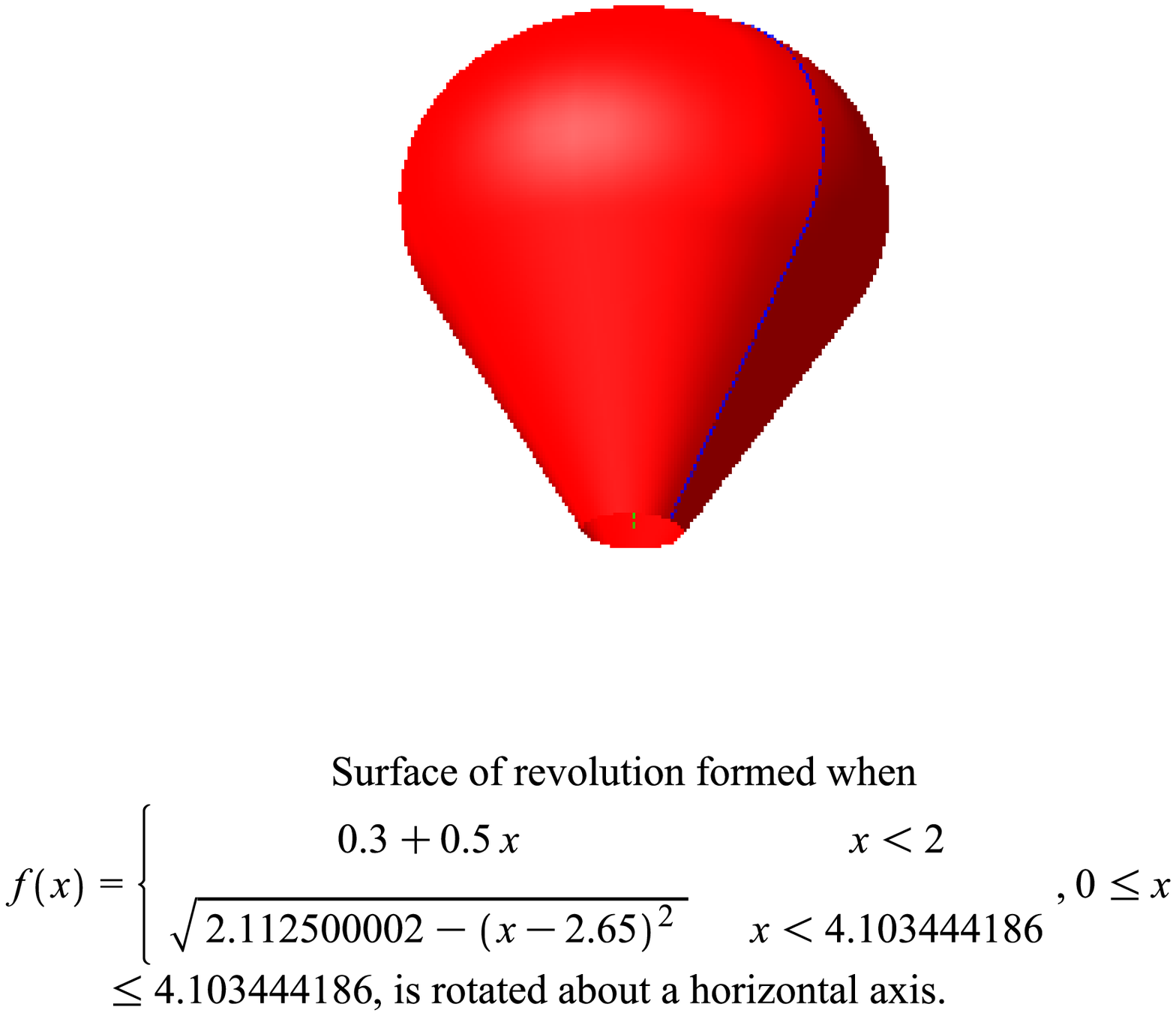}} \caption{\small
Narrow straits formed by a cone-shaped funnel. Axes are rotated $90^o$ about
the $y$-axis.} \label{f:Cone-rev}
\end{figure}
The NET from this type of a domain was calculated in \cite{HHS-PRE} only for
the planar case.

The results of \cite{Ward1}-\cite{PNAS} for small absorbing windows
in a smooth reflecting boundary of a domain $\Omega$ can be
summarized as follows. In the two-dimensional case considered in
\cite{SSH2} the absorbing boundary $\p\Omega_a$ is a small window in
the smooth boundary $\p\Omega$ that is otherwise reflecting to
Brownian trajectories. The MFPT from $\x\in\Omega$ to the absorbing
boundary $\p\Omega_a$, denoted $\bar\tau_{\small\x\to\p\Omega_a}$,
is the NET from the domain $\Omega$ to the small window $\p\Omega_a$
(of length $a$), such that $\eps=\pi|\p\Omega_a|/|\p\Omega|=\pi
a/|\p\Omega|\ll1$ (this corrects the definition in \cite{SSH2}).
Because the singularity of Neumann's function in the plane is
logarithmic the MFPT is given by
\begin{align}
\bar\tau_{\small\x\to\p\Omega_a}=\frac{|\Omega|}{\pi D}\ln
 \frac{|\p\Omega|}{\pi|\p\Omega_a|}+O(1)\hspace{0.5em}\mbox{for}\ \x\in\Omega\
 \mbox{outside a boundary layer near}\ \p\Omega_a.\label{taudisk}
\end{align}

In the three-dimensional case the MFPT to a circular absorbing window
$\p\Omega_a$ of small radius $a$ centered at $\mb{0}$ on the boundary
$\p\Omega$ is given by \cite{Leakage}
 \begin{align}
\bar\tau_{\small\x\to\p\Omega_a}=\frac{|\Omega|}{4aD\left[1+\ds\frac{L(\mb{0})+
N(\mb{0})}{2\pi}\,a\log a+o(a\log a)\right]}\label{tau3D},
 \end{align}
where $L(\mb{0})$ and $N(\mb{0})$ are the principal curvatures of the boundary
at the center of $\p\Omega_a$.

However, the MFPT from a domain to an absorbing interface located at the end of
a funnel, as shown in Figure \ref{f:Canonical-partial-block1}, cannot be
calculated by the methods of \cite{SSH1}-\cite{SSH3}, \cite{Leakage}, because
the contribution of the singular part of Neumann's function to the MFPT in a
composite domain with a funnel or another bottleneck is not necessarily
dominant. Also the method of matched asymptotic expansions, used in
\cite{Ward1}-\cite{Ward3}, \cite{Cheviakov} for calculating the MFPT to the
interface on a smooth boundary, requires major modifications for an interface
at the end of a bottleneck, because the boundary layer problem does not reduce
to the classical electrified disk problem \cite{Jackson}. Altogether different
boundary or internal layers at absorbing windows located at the end of a
cusp-like funnel are needed. The methods used in  \cite{PNAS} and \cite{Berez}
for constructing the MFPT in composite domains of the type shown in Figure
\ref{f:sharp-spine}(right) are made precise here and the new method extends to
domains of the type shown in Figure \ref{f:sharp-spine}(left).

The new results of this paper are as follows. In Section \ref{s:Dire straits}
we prove that the MFPT to the narrow straits formed by a partial block of a
planar domain (see Figures \ref{f:Canonical-partial-block1} and
\ref{f:Partial-block}) is given by
\begin{align}\label{bartaud0}
\bar\tau=\sqrt{\frac{R_c(R_c+r_c)}{2r_c\eps}}\frac{\pi|\,\Omega|}{2D}\left(1+o(1)\right)\hspace{0.5em}\mbox{for}\
\eps\ll|\p\Omega|, R_c, r_c,
\end{align}
where $R_c$ and $r_c$ are the curvatures at the neck and $\eps$ is the width of
the straits. More general cases are also considered. In Section \ref{s:funnel}
we prove that the MFPT in the solid of revolution obtained by rotating the
symmetric domain $\Omega$ in Figure \ref{f:Canonical-partial-block1}(left)
about its axis of symmetry is given by
\begin{align}
\bar\tau=\frac{1}{\sqrt{2}}\left(\frac{R_c}{a}\right)^{3/2} \frac{|\Omega|}{R_c
D}(1+o(1))\hspace{0.5em}\mbox{for}\ a\ll R_c,\label{bartau1/2}
\end{align}
where the radius of the cylindrical neck is $a=\eps/2$. In Section
\ref{s:Diffusion-NET} we consider Brownian motion on a surface of revolution
generated by rotating the curve in Figure
\ref{f:Canonical-partial-block1}(left) about its axis of symmetry.
We use the representation of the generating curve
\begin{align*}
y=r(x),\ \Lambda<x<0\hspace{0.5em}
\end{align*}
where the $x$-axis is horizontal with $x=\Lambda$ at the absorbing end $AB$. We
assume that the parts of the curve that generate the funnel have the form
\begin{align}
r(x)&=O(\sqrt{|x|})\hspace{0.5em}\mbox{near $x=0$}\nonumber\\
r(x)&=a+\frac{(x-\Lambda)^{1+\nu}}{\nu(1+\nu)\ell^{\nu}}(1+o(1))\hspace{0.5em}\mbox{for
$\nu>0$ near $x=\Lambda$},\label{rzL0}
\end{align}
where $a=\frac12\overline{AB}=\eps/2$ is the radius of the gap, and the
constant $\ell$ has dimension of length. For $\nu=1$ the parameter $\ell$ is
the radius of curvature $R_c$ at $x=\Lambda$. We prove that the MFPT from the
head to the absorbing end $AB$ is given by
\begin{align}
\bar\tau\sim\frac{{\cal S}(\Lambda)}{2D}
\frac{\left(\ds\frac{\ell}{(1+\nu)a}\right)^{\nu/1+\nu}\nu^{1/1+\nu}}{\sin\ds\frac{\nu
\pi}{1+\nu}}, \label{u0SD0}
\end{align}
where ${\cal S}$ is the entire unscaled area of the surface. In particular, for $\nu=1$
we get the MFPT
\begin{align}
\bar\tau\sim\frac{{\cal S}}{4D\sqrt{a/2\ell}}.\label{u0Rc}
\end{align}
The case $\nu=0$ corresponds to an absorbing circular cap of small radius $a$ on a closed
surface. For a sphere we get the known result
\begin{equation}
\label{eq:v-sphere0} \bar\tau= \frac{2R^2}{D} \log\frac{\sin\frac
\theta2}{\sin\frac\delta2},
\end{equation}
where $\theta$ is the angle between $\x$ and the south-north axis of the sphere
and $a=R\sin\delta/2$ (see \cite{SSH3}--\cite{Coombs}).

If a right circular cylinder of a small radius $a$ and length $L$ is attached to the
surface at $z=\Lambda$, the NET from the composite surface is given by
\begin{align}
\bar\tau&=\frac{{\cal S}(\Lambda)}{2D}
\frac{\left(\ds\frac{\ell}{(1+\nu)a}\right)^{\nu/1+\nu}\nu^{1/1+\nu}}{\sin\ds\frac{\nu
\pi}{1+\nu}}+\frac{{\cal S}L}{2\pi Da}+\frac{{L}^2}{2D}\hspace{0.5em}\mbox{for}\,
a\ll\ell\label{u0unif0}
\end{align}
(see \cite{Dire-Part-II} for a different derivation). We also find
the NET and the exit probability when there are $N$ absorbing
windows at the ends of narrow necks. These are related to the
principal eigenvalue of the Laplacian in dumbbell-shaped domains
that consists of heads interconnected by narrow necks, which, in
turn, is related to the effective diffusion in such domains. In
Section \ref{s:Connection} we calculate the NET from composite
domains that consist of a head connect by a funnel to a narrow
cylindrical neck and calculate the principal eigenvalue composite
and dumbbell-shaped domains. Finally, in Section
\ref{s:Brownian-needle} we calculate the mean time $\bar\tau$ for a
Brownian needle to turn around in a tightly fitting planar strip.
For a needle of length $l_0$ in a strip of width $l$ it is given by
\begin{align}
\bar\tau=\frac{\pi\left(\ds\frac{\pi}{2}-1\right)}{D_r\sqrt{l_0(l_0-l)}}
\sqrt{\frac{D_X}{D_r}}\left(1+O\left(\sqrt{\frac{l_0-l}{l_0}}\right)\right),\label{turnaround}
\end{align}
where $D_r$ is the rotational diffusion coefficient and $D_X$ is the
translational diffusion coefficient along the needle. We close this
article by providing several applications to cellular biology.

\section{The MFPT to a bottleneck}\label{s:Dire straits}
We consider the NET problem in an asymmetric planar domain, as in
Figure \ref{f:Partial-block} or in an asymmetric version of the
(dimensional) domain $\Omega'$ in Figure
\ref{f:Canonical-partial-block1}. We use the (dimensional)
representation of the boundary curves
\begin{align}
y'=r_\pm(x'),\ \Lambda'<x'<0\hspace{0.5em}\mbox{for the upper and lower parts,
respectively}
\end{align}
where the $x'$-axis is horizontal with $x'=\Lambda'$ at $AB$. We assume that
the parts of the curve that generate the funnel have the form
\begin{align}
r_\pm(x')&=O(\sqrt{|x'|})\hspace{0.5em}\mbox{near $x'=0$}\nonumber\\
r_\pm(x')&=\pm
a'\pm\frac{(x'-\Lambda')^{1+\nu_\pm}}{\nu_\pm(1+\nu_\pm)\ell_\pm^{\nu_\pm}}(1+o(1))\hspace{0.5em}\mbox{for
$\nu_\pm>0$ near $x'=\Lambda'$},\label{rzL}
\end{align}
where $a'=\frac12\overline{AB}=\eps'/2$ is the radius of the gap, and the
constants $\ell_\pm$ have dimension of length. For $\nu_\pm=1$ the parameters
$\ell_\pm$ are the radii of curvature $R^\pm_c$ at $x'=\Lambda'$. To simplify
the conformal mapping, we first rotate the domain by $\pi/2$ clockwise to
assume the shape in Figure \ref{f:Canonical-partial-block1}(left). The rotated
axes are renamed $(x',y')$ as well.

The NET of Brownian motion with diffusion coefficient $D$ from a point
$\x'=(x',y')$ inside the domain $\Omega'$ with reflection at the boundary
$\p\Omega'$, except for an absorbing boundary $\p\Omega_a'$ at the bottom of
the neck, is the solution of the boundary value problem
\begin{align}
D\Delta \bar u(\x')=&\,-1\hspace{0.5em}\mbox{for}\ \x'\in\Omega'\label{PDE'}\\
\frac{\p \bar u(\x')}{\p n}=&\,0\hspace{0.5em}\mbox{for}\ \x'\in\p\Omega'-\p\Omega_a'\nonumber\\
\bar u(\x')=&\,0\hspace{0.5em}\mbox{for}\ \x'\in\p\Omega_a'.\nonumber
\end{align}
We convert to dimensionless variables by setting $\x'=\ell_+\x,\
\Lambda'=\ell_+\Lambda$, the domain $\Omega'$ is mapped into $\Omega$ and we
have (see (\ref{scaling}) below)
 \begin{align}
|\Omega'|=\ell_+^2|\Omega|,\ |\p\Omega'|=\ell_+|\p\Omega|,\
|\p\Omega_a'|=\eps'=\ell_+|\p\Omega_a|=\ell_+\eps.\label{scaling}
\end{align}
Setting $\bar u(\x')=u(\x)$, we write (\ref{PDE'}) as
\begin{align}
\frac{D}{\ell_+^2}\Delta  u(\x)=&\,-1\hspace{0.5em}\mbox{for}\ \x\in\Omega\label{PDE}\\
\frac{\p u(\x)}{\p n}=&\,0\hspace{0.5em}\mbox{for}\ \x\in\p\Omega-\p\Omega_a\nonumber\\
u(\x)=&\,0\hspace{0.5em}\mbox{for}\ \x\in\p\Omega_a.\nonumber
\end{align}

\subsection{Asymptotic analysis}
\begin{figure}
\centering \resizebox{!}{8cm}{\includegraphics{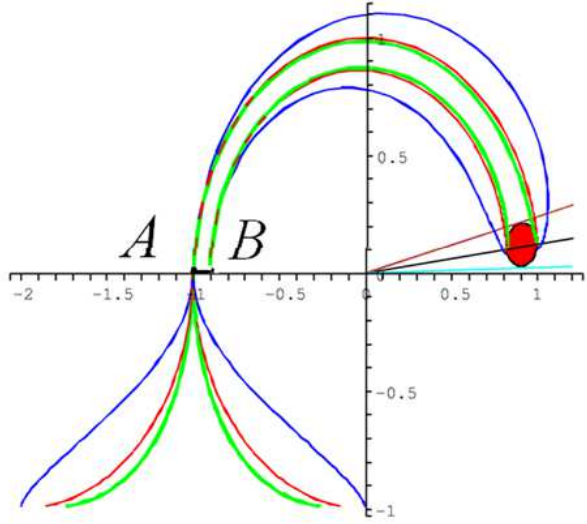}}
\caption{\small The image in $\Omega_w=w(\Omega)$ of the necks $AB$ in Figure
\ref{f:Canonical-partial-block1} under the conformal mapping (\ref{w}). The
funnel in Figure \ref{f:Canonical-partial-block1} (shown inverted in the third
quadrant) is mapped onto the ring enclosed between the like-colored arcs and
the large disk is mapped onto small red disk. The short black segment $AB$ (of
length $\eps$) is mapped onto the thick black segment $\AB$ (of length
$2\sqrt{\eps}+O(\eps)$).} \label{f:Rod-domain}
\end{figure}

First, we consider the case $\nu_\pm=1$, $\ell_+=R_c$, and $l_-=r_c$, radius
$1$, and $A$ has dimensionless radius $r_c/R_c$. This case can represent a
partial block described in Figure \ref{f:Partial-block}. With the scaling
(\ref{scaling}) the bounding circle $B$ has dimensionless We construct an
asymptotic solution for small gap $\eps$ by first mapping the domain $\Omega$
in Figure \ref{f:Canonical-partial-block1}(left) conformally into its image
under the M\"obius transformation of the two bounding circles (thin line) into
concentric circles. To this end we move the origin of the complex plane to the
center of the right lower circle and set
\begin{align}
w=w(z)=\frac{z-\alpha}{1-\alpha z},\label{w}
\end{align}
where
\begin{align}
\alpha=&-\frac{2\eps
R_c+2R_c+\eps^2R_c+2r_c\eps+2r_c}{2(\eps R_c+r_c+R_c)}\nonumber\\
&\pm\frac{\sqrt{\eps(8R_cr_c+4\eps R_c^2+12\eps R_cr_c+4\eps^2R_c^2+8r_c^2+4\eps^2R_cr_c+\eps^3R_c^2+4\eps r_c^2)}}{2(\eps R_c+r_c+R_c)}\nonumber\\
=&-1\pm\sqrt{\frac{2r_c\eps}{R_c+r_c}}+O(\eps),\label{alpha}
\end{align}
which maps the right lower circle into itself and $\Omega$ is mapped onto the
domain $\Omega_w=w(\Omega)$ in Figure \ref{f:Rod-domain}.
The straits in Figure \ref{f:Canonical-partial-block1}(left) are mapped onto
the ring enclosed between the like-colored arcs and the large disk is mapped
onto the small red disk. The radius of the small red disk and the elevation of
its center above the real axis are $O(\sqrt{\eps})$. The short black segment of
length $\eps$ in Figure \ref{f:Canonical-partial-block1} is mapped onto a
segment of length $2\sqrt{\eps}+O(\eps)$.

Setting $u(z)=v(w)$ and $\tilde\eps=2r_c\eps/(R_c+r_c)$, the system (\ref{PDE})
is converted to
\begin{align}
\Delta_wv(w)=&\,-\frac{\ell_+^2}{D|w'(z)|^2}=-\frac{(4\tilde\eps+O(\tilde\eps^{3/2}))\ell_+^2
}{D|w(1-\sqrt{\tilde\eps})-1+O(\tilde\eps)|^4}\hspace{0.5em}
\mbox{for}\ w\in\Omega_w\label{PDEw}\\
\frac{\p v(w)}{\p n}=&\,0\hspace{0.5em}\mbox{for}\ w\in\p\Omega_w-\p\Omega_{w,a}\nonumber\\
v(w)=&\,0\hspace{0.5em}\mbox{for}\ w\in\p\Omega_{w,a}.\nonumber
\end{align}
The MFPT is bounded above and below by that from the inverse image of a
circular ring cut by lines through the origin, tangent to the red disk at polar
angles $\theta=c_1\sqrt{\tilde\eps}$ (brown) and $\theta=c_2\sqrt{\tilde\eps}$
(cyan) for some positive constants $c_1,c_2$, independent of $\tilde\eps$.
Therefore the MFPT from $\Omega$ equals that from the inverse image of a ring
cut by an intermediate angle $\theta=c\sqrt{\tilde\eps}$ (black).

The asymptotic analysis of (\ref{PDEw}) begins with the observation that the
solution of the boundary value problem (\ref{PDEw}) is to leading order
independent of the radial variable in polar coordinates $w=re^{i\theta}$.
Fixing $r=1$, we impose the reflecting boundary condition at
$\theta=c\sqrt{\tilde\eps}$, where $c=O(1)$ is a constant independent of
$\tilde\eps$ to leading order, and the absorbing condition at $\theta=\pi$. The
outer solution, obtained by a regular expansion of $v(e^{i\theta})$, is given
by
\begin{align}
v_0(e^{i\theta})=A(\theta-\pi),\label{v_0}
\end{align}
where $A$ is yet an undetermined constant. It follows that
\begin{align}
\left.\frac{\p v_0(e^{i\theta})}{\p\theta}\right|_{\theta=\pi}=-A.\label{-A}
\end{align}
To determine $A$, we integrate (\ref{PDEw}) over the domain to obtain at the
leading order
\begin{align}
\left.2\sqrt{\tilde\eps}\frac{\p
v_0(e^{i\theta})}{\p\theta}\right|_{\theta=\pi}=-2\sqrt{\tilde\eps}A\sim-\frac{|\Omega'|}{D},\label{solvabilityA}
\end{align}
hence
\begin{align}
A\sim\frac{|\Omega'|}{2D\sqrt{\tilde\eps}}.
\end{align}
Now (\ref{v_0}) gives for $\theta=c\sqrt{\tilde\eps}$ the leading order
approximation
\begin{align}
\bar\tau\sim A\pi=\frac{\pi|\Omega'|}{D\sqrt{\tilde\eps}}.\label{LOAPP}
\end{align}
The following more explicit analysis was briefly summarized in
\cite{HHS-PRE} for the symmetric case $\nu_\pm=1, R_c=r_c$ and is
explicitly given here for completeness. The leading order
approximation is obtained by an explicit integration of (\ref{PDEw})
with respect to $\theta$,
\begin{align}
v\left(e^{i\theta}\right)=\frac{4\ell_+^2\tilde\eps}{D}\int\limits_{\ds\theta}^{\ds\pi}d\varphi\int\limits_{\ds
c\sqrt{\tilde\eps}}^{\ds\varphi} \frac{d\eta}
{|e^{i\eta}-1-e^{i\eta}\sqrt{\tilde\eps}|^4}, \label{Solution01}
\end{align}
so that
\begin{align}
v\left(e^{ic\sqrt{\tilde\eps}}\right)=&\,\frac{4\ell_+^2\tilde\eps}{D}\int\limits_{\ds
c\sqrt{\tilde\eps}}^{\ds\pi}d\varphi\int\limits_{\ds\varphi}^{\ds\pi}\frac{d\eta}
{|e^{i\eta}-1-e^{i\eta}\sqrt{\tilde\eps}|^4}\nonumber\\
=&\,\frac{4\ell_+^2\tilde\eps}{D}\int\limits_{\ds
c\sqrt{\tilde\eps}}^{\ds\pi}\frac{(\pi-\eta)\,d\eta}
{|e^{i\eta}-1-e^{i\eta}\sqrt{\tilde\eps}|^4}.
\end{align}
First, we evaluate asymptotically the integral
\begin{align}
\frac{\ell_+^2\tilde\eps}{D}\int\limits_{\ds
c\sqrt{\tilde\eps}}^{\ds\pi}\frac{\eta\,d\eta}
{|e^{i\eta}-1-e^{i\eta}\sqrt{\tilde\eps}|^4}
\end{align}
by setting $\eta=\sqrt{\tilde\eps}\zeta$ and noting that
\begin{align}
\left|\frac{e^{i\zeta\sqrt{\tilde\eps}}-1}{i\zeta\sqrt{\tilde\eps}}-1\right|=
\left|\frac{-2\sin^2\frac{\zeta\sqrt{\tilde\eps}}{2}}{i\zeta\sqrt{\tilde\eps}}+\frac{\sin\zeta\sqrt{\tilde\eps}}{\zeta\tilde\eps}-1\right|=
O(\zeta\sqrt{\tilde\eps})\hspace{0.5em}\mbox{for all}\ \eta,\tilde\eps>0.
\end{align}
It follows that
\begin{align}
\frac{4\ell_+^2\tilde\eps}{D}\int\limits_{\ds
c\sqrt{\tilde\eps}}^{\ds\pi}\frac{\eta\,d\eta}
{|e^{i\eta}-1-e^{i\eta}\sqrt{\tilde\eps}|^4}=\frac{4\ell_+^2}{D}\int\limits_{\ds
c}^{\ds\pi/\sqrt{\tilde\eps}}
\frac{\zeta\,d\zeta}{|1+\zeta^2+O(\tilde\eps\zeta^2)|^2}=\frac{4}{D(c+1)}\left(1+O(\sqrt{\tilde\eps})\right).
\end{align}
Similarly, we obtain that
\begin{align}
\frac{4\tilde\eps}{D}\int\limits_{\ds c\sqrt{\tilde\eps}}^{\ds\pi}\frac{d\eta}
{|e^{i\eta}-1-e^{i\eta}\sqrt{\tilde\eps}|^4}=\frac{4}{D\sqrt{\tilde\eps}}\int\limits_{\ds
c}^{\ds\pi/\sqrt{\tilde\eps}}
\frac{d\zeta}{|1+\zeta^2+O(\tilde\eps\zeta^2)|^2}=\frac{C}{D\sqrt{\tilde\eps}}\left(1+O(\sqrt{\tilde\eps})\right),
\end{align}
where $C=O(1)$ is a constant, so that
\begin{align}
v\left(e^{ic\sqrt{\tilde\eps}}\right)=\frac{4\ell_+^2\pi
C}{D\sqrt{\tilde\eps}}\left(1+O(\sqrt{\tilde\eps})\right).
\end{align}
To determine the value of the constant $C$, we note that (\ref{Solution01})
implies that
\begin{align}\label{dvdn}
\left.\frac{\p v\left(e^{i\theta}\right)}{\p
n}\right|_{\p\Omega_{w,a}}=\left.\frac{\p
v}{\p\theta}\right|_{\theta=\pi}=-\frac{4\ell_+^2\tilde\eps}{D}\int\limits_{\ds
c\sqrt{\tilde\eps}}^{\ds\pi}\frac{d\eta}
{|e^{i\eta}-1-e^{i\eta}\sqrt{\tilde\eps}|^4}=-\frac{4\ell_+^2C}{D\sqrt{\tilde\eps}}\left(1+O(\sqrt{\tilde\eps})\right)
\end{align}
and the integration of (\ref{PDEw}) over $\Omega_w$ gives
\begin{align}\label{epsdvdn}
2\sqrt{\tilde\eps}\left.\frac{\p v\left(e^{i\theta}\right)}{\p
n}\right|_{\p\Omega_{w,a}}=-\frac{\ell_+^2|\Omega|}{D}.
\end{align}
Now, (\ref{dvdn}) and (\ref{epsdvdn}) imply that $4C=|\Omega|/2$, so that the
MFPT to the straits, $\bar\tau$, is
\begin{align}\label{bartau1}
\bar\tau=\frac{\ell_+^2\pi|\,\Omega|}{2D\sqrt{\tilde\eps}}\left(1+o(1)\right)=\frac{\pi|\,\Omega'|}{2D\sqrt{\tilde\eps}}\left(1+o(1)\right)\hspace{0.5em}\mbox{for}\
\tilde\eps\ll|\p\Omega|,\ell_+,
\end{align}
which is (\ref{LOAPP}). In dimensional units (\ref{bartau1}) becomes
\begin{align}\label{bartaud}
\bar\tau=\sqrt{\frac{R_c(R_c+r_c)}{2r_c\eps'}}\frac{\pi|\,\Omega'|}{2D}\left(1+o(1)\right)\hspace{0.5em}\mbox{for}\
\eps'\ll|\p\Omega'|,R_c,r_c.
\end{align}
In the symmetric case $R_c=r_c$ (\ref{bartaud}) reduces to the result of
\cite{HHS-PRE}
\begin{align}\label{bartaudS}
\bar\tau=\frac{\pi|\,\Omega'|}{2D\sqrt{\eps'/R_c}}\left(1+o(1)\right)\hspace{0.5em}\mbox{for}\
\eps'\ll|\p\Omega'|,R_c.
\end{align}
Next, we consider for simplicity the symmetric case $\nu_+=\nu_->1$, so
$R_c=r_c=\infty$. After scaling the boundary value problem (\ref{PDE'}) with
(\ref{scaling}), we can choose the bounding circles at $A$ and $B$ to have
radius $1$ and repeat the above analysis in the domain $\Omega_w$ enclosed by
the green curves, shown in Figure (\ref{f:Rod-domain}). The result
(\ref{bartaudS}) becomes
\begin{align}\label{bartaudl}
\bar\tau=\frac{\pi|\,\Omega'|}{2D\sqrt{\eps'/\ell_+}}\left(1+o(1)\right)\hspace{0.5em}\mbox{for}\
\eps'\ll|\p\Omega'|,\ell_+.
\end{align}
\subsection{Exit from several bottlenecks}\label{ss:exitprob}
In case of exit through any one of $N$ well-separated necks with dimensionless
curvature parameters $l_j$ and widths $\tilde\eps_j$, we construct the outer
solution (\ref{v_0}) at any one of the $N$ absorbing windows so that (\ref{-A})
holds at each window. The integration of (\ref{PDEw}) over $\Omega$ gives the
following analog of (\ref{solvabilityA}),
\begin{align}
\sum\limits_{j=1}^N\left.2\sqrt{\tilde\eps_j}\frac{\p
v_0(e^{i\theta})}{\p\theta}\right|_{\theta=\pi}=-2\sum\limits_{j=1}^N\sqrt{\tilde\eps_j}A
\sim-\frac{|\Omega'|}{D},\label{solvabilityAj}
\end{align}
hence
\begin{align}
A\sim\frac{|\Omega'|}{2D\sum_{j=1}^N\sqrt{\tilde\eps_j}}.
\end{align}
Equation (\ref{bartau1}) is then generalized to
\begin{align}\label{bartaui}
\bar\tau=\frac{\pi|\Omega'|}{2D\sum_{j=1}^N\sqrt{\eps_j'/\ell_j}}\left(1+o(1)\right)\hspace{0.5em}\mbox{for}\
\eps_j'/\ell_j\ll|\p\Omega|.
\end{align}
Equations (\ref{bartaud})-(\ref{bartaudl}) are generalized in a similar manner.

To calculate the exit probability through any one of the $N$ necks, we apply
the transformation (\ref{w}) separately for each bottleneck at the absorbing
images $\p\Omega_{w,a_1},\ldots,\p\Omega_{w,a_N}$ to obtain images
$\Omega_{w_j}$ for $j=1,2,\ldots,N$. Then the probability of exiting through
$\p\Omega_{w,a_i}$ is the solution of the mixed boundary value problem
\begin{align}
\Delta_wv(w)=&\,0\hspace{0.5em}
\mbox{for}\ w\in\Omega_{w_i}\label{PDEwi}\\
\frac{\p v(w)}{\p n}=&\,0\hspace{0.5em}\mbox{for}\ w\in\p\Omega_{w_i}-\bigcup\limits_{i=1}^N\p\Omega_{w,a_i}\nonumber\\
v(w)=&\,1\hspace{0.5em}\mbox{for}\ w\in\p\Omega_{w,a_i}\nonumber\\
v(w)=&\,0\hspace{0.5em}\mbox{for}\ w\in\p\Omega_{w,a_j},\ j\neq i.\nonumber
\end{align}
The outer solution, which is the exit probability through window
$\p\Omega_{w,i},$ is an unknown constant $p_i$. We construct boundary layers at
each absorbing boundary $\p\Omega_{w,a_j}$ for $j\neq i$ by solving the
boundary value problem in $\Omega_{w_j}$, which is of the type shown in Figure
\ref{f:Rod-domain} with a neck of width $\eps_j$. In each case the boundary
layer is a linear function
\begin{align}
v_j(\theta)=\delta_{i,j}-A_j(\theta-\pi)\hspace{0.5em}\mbox{for all}\ j,
\end{align}
such that
\begin{align}
v_j(0)\sim\delta_{i,j}+A_j\pi=p_i\hspace{0.5em}\mbox{for all}\ j.\label{C}
\end{align}
To determine the value of the constant $p_i$, we note that
\begin{align}
\left.\frac{\p v\left(e^{i\theta}\right)}{\p
n}\right|_{\p\Omega_{w,a}}=\left.\frac{\p
v_j(\theta)}{\p\theta}\right|_{\theta=\pi}=-A_j,
\end{align}
so the integration of (\ref{PDEwi}) over $\Omega_{w_i}$ gives
\begin{align}
\sum\limits_{j=1}^NA_{j}|\p\Omega_{w,a_j}|=\sum\limits_{j=1}^N2A_{j}\sqrt{\tilde\eps_j}=0.\label{sum0}
\end{align}
The $N+1$ equations (\ref{C}) and (\ref{sum0}) for the unknowns
$p_i,A_1,\ldots,A_N$ give the exit probability from an interior point in the
planar case as
\begin{align}
p_i=\frac{\sqrt{\eps'/\ell_i}}{\sum_{j=1}^N\sqrt{\eps_j'/\ell_j}}.
\end{align}
\section{The NET in a solid funnel-shaped domain}\label{s:funnel}
We consider now the NET problem in the solid of revolution obtained by rotating
the symmetric domain $\Omega'$ in Figure \ref{f:Canonical-partial-block1}(left)
about its axis of symmetry. The absorbing end of the neck becomes a circular
disk of radius $a'=\eps'/2$. Due to cylindrical symmetry of the mixed boundary
value problem (\ref{PDE}) the MFPT in cylindrical coordinates centered on the
axis of symmetry is independent of the angle. It follows that with the scaling
(\ref{scaling}) the boundary value problem (\ref{PDE}) in the scaled spatial
domain $\Omega$ can be written in cylindrical coordinates as
\begin{align}
\Delta u=\frac{\p^2u}{\p r^2} + \frac{1}{r}\frac{\p u}{\p r} +\frac{\p^2u}{\p
z^2}=-\frac{\ell_+^2}{D}\label{2DPDE}.
\end{align}
Equation (\ref{2DPDE}) can be considered as a two-dimensional problem in the
planar cross section by a plane through the axis of symmetry of $\Omega$ in the
$(r,z)$ plane. Here $r$ is the distance to the axis of symmetry of $\Omega$,
the $z$ axis is perpendicular to that axis and the origin is inside the cross
section of $\Omega$, at the intersection of the axis with the tangent to the
osculating circle to the cross section at the gap. Setting $u_1=u r^{1/2}$, the
MFPT equation (\ref{2DPDE}) takes the form
\begin{align}
\frac{\p^2 u_1(r,z)}{\p r^2}+\frac{\p^2 u_1(r,z)}{\p
z^2}=-\frac{\ell_+^2}{D}\left(r^{1/2}+\frac{u_1(r,z)}{4r^2}\right)\label{ueq}
\end{align}
in the cross section, with mixed Neumann-Dirichlet boundary conditions, as in
the planar case. We assume that in dimensionless variables
$\overline{\AB}=\eps\ll1<|\Omega|^{1/3}$, so the funnel is a narrow passage.
The transformation to the rotated and translated coordinates is given by
$\tilde r=r-1-\eps/2,\ \tilde z=-z+1$. Setting $u_1(r,z)=\tilde u(\tilde
r,\tilde z)$, equation (\ref{ueq}) becomes
\begin{align}
\frac{\p^2 \tilde u(\tilde r,\tilde z)}{\p \tilde r^2}+\frac{\p^2 \tilde
u(\tilde r,\tilde z)}{\p \tilde z^2}=-\frac{\ell_+^2}{D}\left(\left(\tilde
r+1+\frac{\eps}{2}\right)^{1/2}-\frac{\tilde u(\tilde r,\tilde
z)}{4\left(\tilde r+1+\ds\frac{\eps}{2}\right)^2}\right).\label{ueqtilde}
\end{align}
\subsection{Asymptotic solution}\label{ss:Asymptotic}
\begin{figure}
\centering \resizebox{!}{8cm}{\includegraphics{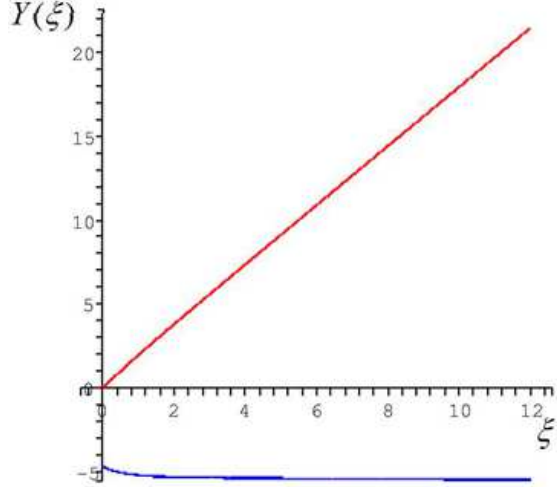}}
\caption{\small Two linearly independent solutions of (\ref{BLEQ}). The
linearly growing solution $Y_1(\xi)$ (red) satisfies the initial conditions
$Y_1(0)=0,\,Y_1'(0)=2$. The asymptotically constant solution $Y_2(\xi)$ (blue)
satisfies the initial conditions $Y_2(0)=-4.7,\,Y_2'(0)=-1$. The asymptotic
value is $Y_2(\infty)\approx-5$.} \label{f:Boundary-layers}
\end{figure}

We construct an asymptotic solution for small gap $\eps$ by first mapping the
cross section in the $(r,z)$-plane conformally into its image under the
M\"obius transformation (\ref{w}),
\begin{align}
w(\zeta)=\rho e^{i\eta}=\frac{\zeta-\alpha}{1-\alpha \zeta},\label{wzeta}
\end{align}
where $\alpha$ is given in (\ref{alpha}) for the symmetric case $R_c=r_c=1$.
Setting $\tilde u(\zeta)=v(w)$, equation (\ref{ueqtilde}) becomes
\begin{align}
\Delta_wv(w)=
&\,\frac{\ell_+^2}{D|w'(\zeta)|^2}\left(-{\left|{\cal{R}}\mbox{e}\frac{w+\alpha}{1+\alpha
w}+1+\ds\frac{\eps}{2}\right|}^{1/2}-\frac{v}{4\ds\left|{\cal{R}}\mbox{e}\frac{w+\alpha}{1+\alpha
w}+1+\ds\frac{\eps}{2}\right|^2}\right).
\end{align}
Because the normalized head of Figure \ref{f:Canonical-partial-block1}(left) is
mapped into the narrow hot dog--shaped region in Figure  \ref{f:Rod-domain} of
width $\sqrt{\eps}$ at $\rho=1$, we approximate
\begin{align}
w=e^{i\eta}+O(\sqrt{\eps}),\quad\left|\frac{w+\alpha}{1+\alpha
w}\right|=1+O(\sqrt{\eps}).\label{simplification}
\end{align}
We also have
\begin{align}
w'(\zeta)=&\,\frac{(1+\alpha w)^2}{\alpha^2-1}\label{w'wp}\\
|w'(\zeta)|^2=&\,\left|\frac{(1+w\alpha)^2}{1-\alpha^2}\right|^2=\frac{|1-w+\sqrt{\eps}\,w|^4}{4\eps}
(1+O(\sqrt{\eps}))\label{w'2w},
 \end{align}
so that (\ref{ueq}) reduces to
\begin{align}
\Delta_wv=
&\,-\frac{\ell_+^2}{D}\frac{4\eps(1+O(\sqrt{\eps}))}{|1-w+\sqrt{\eps}\,w|^4}
\left(\sqrt{2}+\frac{1}{16}v\right),
\end{align}
or equivalently,
\begin{align}
v''+\frac{\eps}{4|e^{i\eta}-1-e^{i\eta}\sqrt{\eps}|^4}v=
\frac{\ell_+^2}{D}\frac{4\sqrt{2}\eps}
{|e^{i\eta}-1-e^{i\eta}\sqrt{\eps}|^4}\left(1+O(\sqrt{\eps})\right).
\end{align}
Setting $v=\ell_+^2(y-16\sqrt{2})/D$, we obtain the leading order equation
\begin{align}
y''(\eta)+\frac{\eps}{4|e^{i\eta}-1-e^{i\eta}\sqrt{\eps}|^4}y(\eta)=0.
\end{align}
The boundary conditions are
\begin{align}
y'(c\sqrt{\eps})=0,\quad y(\pi)=16\sqrt{2}.\label{BCs}
\end{align}
The outer solution is the linear function
\begin{align}
y_{\mbox{\scriptsize outer}}(\eta)=M\eta+N,
\end{align}
where $M$ and $N$ are yet undetermined constants. The absorbing boundary
condition in (\ref{BCs}) gives
\begin{align}
y_{\mbox{\scriptsize outer}}(\pi)=M\pi+N=16\sqrt{2}.
\end{align}
A boundary layer correction is needed to satisfy the boundary conditions at the
reflecting boundary at $\eta=c\sqrt{\eps}$. To resolve the boundary layer at
$\eta=c\sqrt{\eps}$, we set $\eta=\sqrt{\eps}\xi$ and expand
 \begin{align*}
\frac{\eps^2}{|e^{i\eta}-1-e^{i\eta}\sqrt{\eps}|^4}
=\frac{1}{(1+\xi^2)^2}+O(\sqrt{\eps}).
\end{align*}
Writing $y_{\mbox{\scriptsize bl}}(\eta)=Y(\xi)$, we obtain to leading order
the boundary layer equation
 \begin{align}
 Y''(\xi)+\frac{1}{4(1+\xi^2)^2}Y(\xi)=0,\label{BLEQ}
 \end{align}
which has two linearly independent solutions, $Y_1(\xi)$ and $Y_2(\xi)$ that
are linear functions for sufficiently large $\xi$. Initial conditions for
$Y_1(\xi)$ and $Y_2(\xi)$ can be chosen so that $Y_2(\xi)\to const$ as
$\xi\to\infty$ (e.g., $Y_2(0)=-4.7,\ Y_2'(0)=-1$, see Figure
\ref{f:Boundary-layers}).
Setting
\begin{align}
y_{\mbox{\scriptsize bl}}(\eta)=AY_1\left(\frac{\eta}{\sqrt{\eps}}\right)
+BY_2\left(\frac{\eta}{\sqrt{\eps}}\right),
\end{align}
where $A$ and $B$ are constants to be determined, we seek a uniform
approximation to $y(\eta)$ in the form $y_{\mbox{\scriptsize
unif}}(\eta)=y_{\mbox{\scriptsize outer}}(\eta)+y_{\mbox{\scriptsize
bl}}(\eta)$. The matching condition is that $AY_1\left(\eta/\sqrt{\eps}\right)+
+BY_1\left(\eta/\sqrt{\eps}\right)$ remains bounded as $\xi\to\infty$, which
implies $A=0$. It follows that at the absorbing boundary $\eta=\pi$ we have
\begin{align}
y_{\mbox{\scriptsize unif}}(\pi)=&\,M\pi+\beta-5B=16\sqrt{2}\label{absBC}
\\
y_{\mbox{\scriptsize unif}}'(\pi)=&\,M.\nonumber
\end{align}
At the reflecting boundary we have to leading order
\begin{align}
y_{\mbox{\scriptsize unif}}'(c\sqrt{\eps})=&\,y_{\mbox{\scriptsize
outer}}'(c\sqrt{\eps})+y_{\mbox{\scriptsize bl}}'(c\sqrt{\eps}) =
M+B\frac{Y_2'(c)}{\sqrt{\eps}}=0,\label{ypu}
\end{align}
which gives
\begin{align}
B=-\frac{M\sqrt{\eps}}{Y_2'(c)},\quad
N=16\sqrt{2}-\frac{5M\sqrt{\eps}}{Y_2'(c)}-M\pi.
\end{align}
The uniform approximation to $v(w)$ is given by
\begin{align}
v_{\mbox{\scriptsize unif}}(\rho
e^{i\eta})=M\left(\eta-\pi-\frac{5\sqrt{\eps}}{Y_2'(c)}\right),\label{vunifp}
\end{align}
so that using (\ref{w'wp}), we obtain from (\ref{vunifp})
\begin{align}
\left.\frac{\p u}{\p n}\right|_{\zeta\in\p\Omega_a}=\left.\frac{\p
v(\rho e^{i\eta})}{\p
\eta}\right|_{\eta=\pi}w'(\zeta)\Big|_{\zeta=-1}=\frac{2M}{\sqrt{\eps}}(1+O(\sqrt{\eps})).\label{dudnp}
\end{align}
To determine the value of $M$, we integrate (\ref{PDE}) over
$\Omega$, use (\ref{dudnp}), and the fact that
\begin{align}
\int\limits_{\p\Omega_a}\,dS=\frac{\pi\eps^2}{4},\label{Compat1}
\end{align}
to obtain $M=-2\ell_+^2|\Omega|/D\pi\eps^{3/2}$. Now (\ref{vunifp})
gives the MFPT at any point $\x$ in the head as
\begin{align}
\bar\tau=u(\x)\sim v\left(\rho
e^{c\sqrt{\eps}}\right)\sim2\eps^{-3/2}\frac{\ell_+^2|\Omega|}{D}=2\eps^{-3/2}
\frac{|\Omega'|}{\ell _+D}\hspace{0.5em}\mbox{for}\ \eps\ll1.\label{bartau3/2}
\end{align}
The dimensional radius of the absorbing end of the funnel is $a'=\ell_+\eps/2$
(see (\ref{scaling})), so (\ref{bartau3/2}) can be written in physical units as
\begin{align}
\bar\tau=\frac{1}{\sqrt{2}}\left(\frac{\ell_+}{a'}\right)^{3/2} \frac{V}{\ell_+
D}(1+o(1))\hspace{0.5em}\mbox{for}\ a'\ll \ell_+,\label{bartau1/2f}
\end{align}
where $V=|\Omega'|$ is the volume of the domain.

\subsection{Exit from several bottlenecks}
The generalization of (\ref{bartau1/2f}) to exit through $N$
well-separated necks is found by noting that (\ref{Compat1}) becomes
\begin{align}
\int\limits_{\p\Omega_a}\,dS=\sum\limits_{j=1}^N\frac{\pi\eps_j^2}{4},\label{AreaN}
\end{align}
and the integration of (\ref{PDE'}) over $\Omega'$ gives the compatibility
condition (dimensional)
\begin{align}
\int_{\p\Omega'}\frac{\p u(\x')}{\p
n'}\,dS'=M\sum\limits_{j=1}^N\frac{\ell_j\pi\eps_j^2}{4\sqrt{\eps_j}}=-\frac{|\Omega'|}{D}
\end{align}
which determines
\begin{align}
M=-\frac{4|\Omega'|}{D\sum_{j=1}^N\ell_j\pi\eps_j^{3/2}}.
\end{align}
Hence, using the dimensional $a_j'=\ell_j\eps_j/2$, we obtain
\begin{align}
\bar\tau=-M\pi=\frac{1}{\sqrt{2}}\,\frac{|\Omega'|}{D\sum_{j=1}^N\ell_j\ds{\left(\frac{a_j'}{\ell_j}\right)^{3/2}}}
\label{bartauN}.
\end{align}
To calculate the exit probability from one of $N$ necks, we note that the
boundary layer function is to leading order linear, as in Section
\ref{ss:exitprob}. Therefore in the three-dimensional case the exit probability
is given by
\begin{align}
p_i=\frac{\eps_i^{3/2}\ell_i}{\sum_{j=1}^N\eps_j^{3/2}\ell_j}=\frac{{a_i'}^{3/2}\ell_i^{-1/2}}{\sum_{j=1}^N{a_j'}^{3/2}\ell_j^{-1/2}}.
\end{align}
\section{Diffusion and NET on a surface of revolution}\label{s:Diffusion-NET}
We consider now Brownian motion on a surface of revolution generated by
rotating the curve in Figure \ref{f:Canonical-partial-block1}(left) about its
axis of symmetry and assume $\nu_+=\nu_-=\nu$ and $\ell_+=\ell_-=\ell$. The
projection of the Brownian motion from the surface to the $z$-axis gives rise
to a drift. The backward Kolmogorov operator \cite{DSP} of the projected
motion, scaled with (\ref{scaling}), is given by
\begin{align}
{\cal
L}^*u(z)=\frac{D}{\ell^2}\left\{\frac{1}{1+{r'}^2(z)}u''(z)+\left[\frac{r'(z)}{r(z)(1+{r'(z)}^2)}-
\frac{r'(z)r''(z)}{(1+{r'(z)}^2)^2}\right]u'(z)\right\}.
\end{align}
The operator ${\cal L}^*$ corresponds to the It\^o equation
\begin{align}
dz=a(z)\,dt+b(z)\,dw,
\end{align}
where the drift $a(z)$ and noise intensity $b(z)$ are given by
\begin{align}
a(z)=\frac{D}{\ell^2}\left\{\frac{r'(z)}{r(z)(1+{r'(z)}^2)}-\frac{r'(z)r''(z)}{(1+{r'(z)}^2)^2}\right\},\quad
b(z)=\sqrt{\frac{2D}{\ell^2(1+{r'}^2(z))}}\label{azbz}
\end{align}
and $w(t)$ is standard Brownian motion on the line. The potential of the drift is
$A(z)=-\int_\Lambda^za(t)\,dt.$

To calculate the MFPT from $z=0$ to the end of the funnel at $z=\Lambda$, we note that
due to rotational symmetry the solution of the Andronov-Pontryagin-Vitt boundary value
problem \cite{DSP} for the MFPT $u(z,\theta)$ on the surface  is independent of $\theta$.
Therefore the problem reduces to
\begin{align}
\frac{1}{r(z)\sqrt{1+{r'}^2(z)}}\frac{\p}{\p
z}\left[\frac{r(z)}{\sqrt{1+{r'}^2(z)}}\frac{\p u(z)}{\p z}\right]&=-\frac{\ell^2}{D}\label{zeq}\\
u'(0)=u(\Lambda)&=0.\nonumber
\end{align}
The MFPT is given by
\begin{align}
u(0)=\frac{\ell^2}{2\pi
D}\int\limits_\Lambda^0\frac{\sqrt{1+{r'}^2(t)}}{r(t)}S(t)\,dt.\label{u0}
\end{align}
where $S(t)$ is the (scaled) area of the surface of revolution from $z=t$ to
$z=0$, given by
\begin{align}
S(t)=2\pi\int\limits_t^0r(s)\sqrt{1+{r'}^2(s)}\,ds.\label{Sarea}
\end{align}
The main contribution to (\ref{u0}) comes from $\Lambda<t<\Lambda+\delta$ for a
sufficiently small $\delta$, such that $\delta\gg a$ (note that the singularity
of $1/r(z)$ near $z=0$ is integrable). Thus (\ref{u0}) and (\ref{Sarea}) give
for $\nu>0$
\begin{align}
\bar\tau=u(0)\sim\frac{\ell^2S(\Lambda)}{2\pi
D}\int\limits_\Lambda^{\Lambda+\delta}\frac{\sqrt{1+{r'}^2(t)}}{r(t)}\,dt\sim\frac{{\cal
S}(\Lambda)}{2D}
\frac{\left(\ds\frac{\ell}{(1+\nu)a}\right)^{\nu/1+\nu}\nu^{1/1+\nu}}{\sin\ds\frac{\nu
\pi}{1+\nu}}, \label{u0SD}
\end{align}
where ${\cal S}={\cal S}(\Lambda)$ is the entire unscaled area of the surface.
In particular, for $\nu=1$ we get the MFPT
\begin{align}
\bar\tau\sim\frac{{\cal S}}{4D\sqrt{a/2\ell}}.\label{u0Rc1}
\end{align}
The case $\nu=0$ corresponds to an absorbing circular cap of a small radius $a$ on a
closed surface. For a sphere the solution of (\ref{zeq}) gives the known result
\begin{equation}
\label{eq:v-sphere} \bar\tau_{\small\x\to\p\Omega_i} = \frac{2R^2}{D}
\log\frac{\sin\frac \theta2}{\sin\frac\delta2},
\end{equation}
where $\theta$ is the angle between $\x$ and the south-north axis of the sphere
and $a=R\sin\delta/2$ (see \cite{SSH3}--\cite{Coombs}).
\begin{figure}
\centering
\resizebox{!}{5cm}{\includegraphics{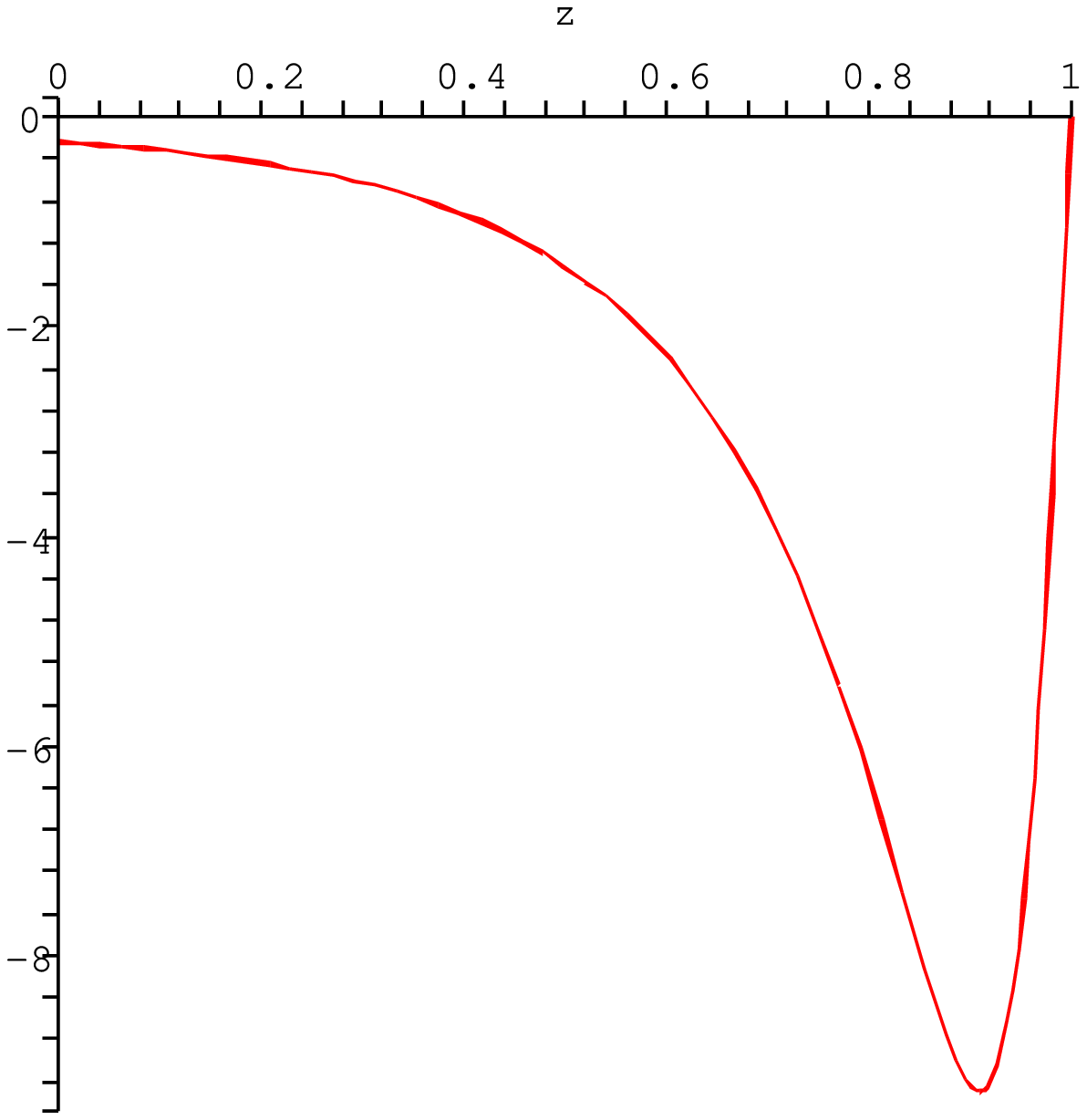}}\resizebox{!}{5cm}{\includegraphics{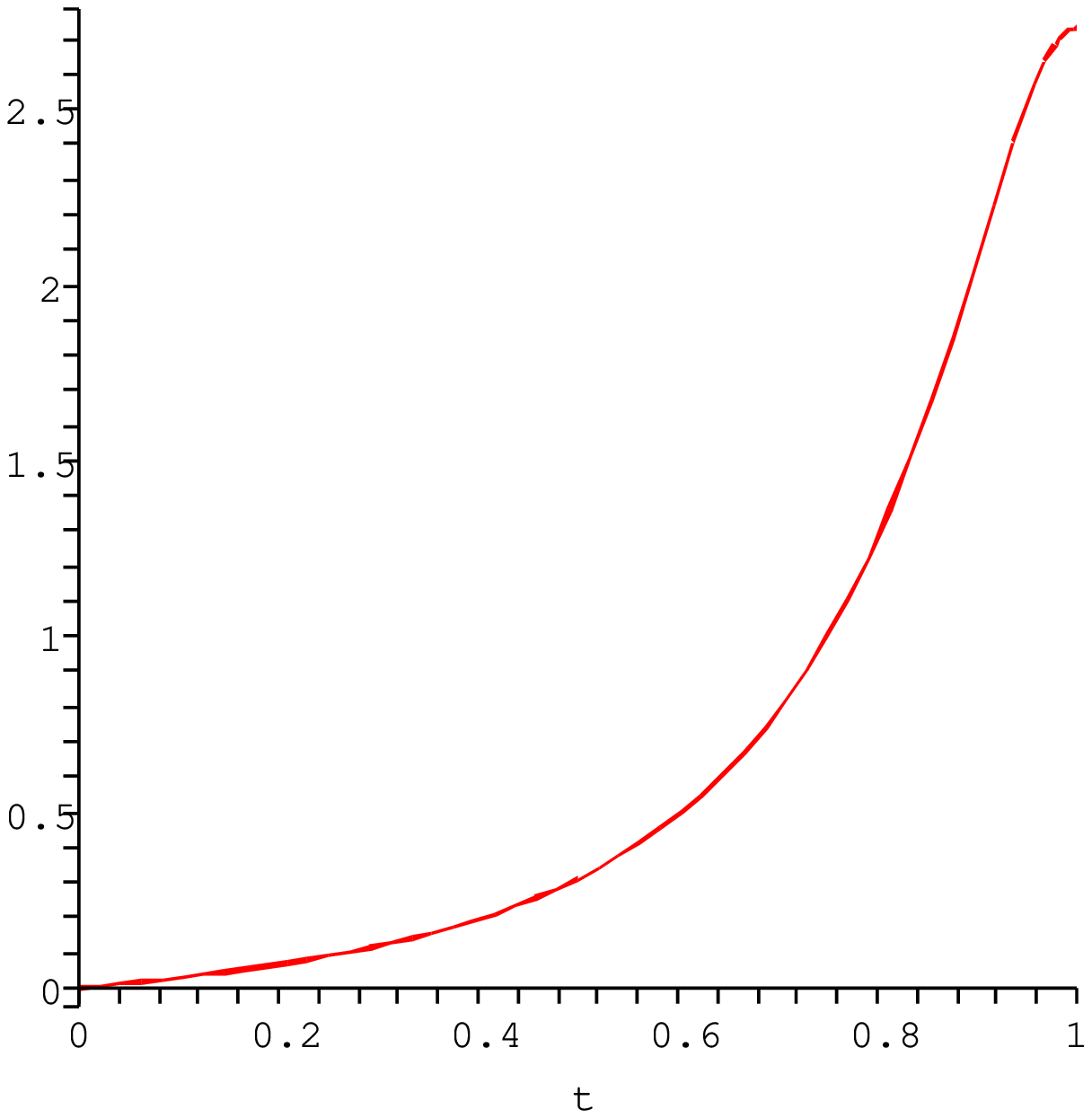}}
\caption{\small The drift $a(z)$ in (\ref{azbz}) (left panel) and its potential
$A(z)$ (right panel) near the cusp. The projection of the Brownian motion on
the axis of symmetry has an effective high barrier in the neck.}
\label{f:Drift-Pot}
\end{figure}
If a right circular cylinder of a small radius $a$ and length $L'=\ell L$ is attached to
the surface at $z=\Lambda$, then the integration in (\ref{u0SD}) extends now to
$\Lambda-L$, giving
\begin{align}
u(0)&\sim \frac{\ell^2S(\Lambda)}{2\pi
D}\int\limits_\Lambda^0\frac{\sqrt{1+{r'}^2(t)}}{r(t)}\,dt+\frac{\ell^2}{2\pi
Da}\int\limits_{\Lambda-L}^\Lambda[S(\Lambda)+2\pi
a(t-\Lambda)]\,dt\nonumber\\
&=\frac{{\cal S}(\Lambda)}{2\pi
D}\int\limits_\Lambda^0\frac{\sqrt{1+{r'}^2(t)}}{r(t)}\,dt+\frac{{\cal
S}(\Lambda)L'}{2\pi Da}+\frac{{L'}^2}{2D},\label{u0unif}
\end{align}
where the integral is given by (\ref{u0SD}), (\ref{u0Rc1}), or
(\ref{eq:v-sphere}) for the various values of $\nu$. Note that while
$\bar\tau$ on the surface depends on the fractional power
$-\nu/(1+\nu)$ of the neck's radius $a$, the power of $a$ in the
three-dimensional case is $-3/2$, as indicated in
(\ref{bartau1/2f}).

The case $\nu=0$ is not the limit of (\ref{u0SD}), because the line
(\ref{rzL0}) blows up. This case corresponds to a conical funnel with an
absorbing circle of small radius $a$ and length $H$  (see Figure
\ref{f:Cone-rev}). We assume that the radius of the other base of the cone,
$b$, is smaller than $a$, but that $b\ll {\cal S}^{1/2}$. The generator of the
cone is the line segment
 \beq
 r(x)=a+C(x-L)\hspace{0.5em}\mbox{for}\ \Lambda-L<x<\Lambda,
 \eeq
where $C$ is the (positive) slope. In this case (\ref{u0unif}) is
replaced by
\begin{align*}
u(0)=&\,\frac{{\cal S}(\Lambda)}{2\pi
D}\int\limits_\Lambda^0\frac{\sqrt{1+{r'}^2(t)}}{r(t)}\,dt+\frac{{\cal
S}(\Lambda)\sqrt{1+C^2}}{2\pi
DC}\log\left(1+\frac{CL'}{a}\right)\\
&\,+\frac{(1+C^2)}{2DC^2}\left[(a+CL')\log\left(1+\frac{CL'}{a}\right)+\frac{1}{2}[(a+CL')^2-a^2]\right],
\end{align*}
which reduces to (\ref{u0unif}) in the limit $CL'\ll a$ and for
$a\ll CL'$ can be simplified to leading order to
\begin{align}
u(0)=&\,\frac{{\cal S}(\Lambda)}{2\pi
D}\int\limits_\Lambda^0\frac{\sqrt{1+{r'}^2(t)}}{r(t)}\,dt+\frac{{\cal
S}(\Lambda)\sqrt{1+C^2}}{2\pi
DC}\log\frac{CL'}{a}\nonumber\\
&\,+\frac{(1+C^2)L'^2}{2D}\log\frac{CL'}{a}+O(1).\label{cone}
\end{align}
Note that the last term in (\ref{cone}) blows up as $a\to0$ while
that in (\ref{u0unif}) does not. This is due to the degeneration of
the NET problem in the cylinder, as noted in \cite{SSH1}.

\section{The principal eigenvalue in domains with bottlenecks}\label{s:Connection}
The narrow escape time is related to the leading eigenvalues of the
Neumann or mixed Neumann-Dirichlet problem for the Laplace equation
in domains that consists of compartments and narrow necks. In
domains that consists of compartments interconnected by narrow necks
the MFPT from one compartment to the other, as defined in
\cite{equilibrium}, is to leading order (in the limit of shrinking
neck) independent of the initial point of the escaping trajectory
and is twice the MFPT from the compartment to the narrowest passage
in the bottleneck (e.g., the interval $\AB$ in Figure
\ref{f:dumbbell}). Indeed, the reciprocal of this MFPT is to leading
order the rate at which trajectories reach the bottleneck from the
first compartment, so the reciprocal of the MFPT is the lowest
eigenvalue of the mixed Neumann-Dirichlet boundary value problem in
the first compartment with Dirichlet conditions on the cross section
of the neck.

There is a spectral gap of order 1 from the smallest eigenvalue to
the next one. It follows that long transition times of Brownian
trajectories between compartments connected by bottlenecks are
exponentially distributed and therefore the leading eigenvalues of
Neumann's problem for the Laplace equation in a domain that consists
of compartments interconnected by narrow necks are to leading order
the eigenvalues of a Markov chain with transition rates that are the
reciprocals of the MFPTs through the narrow necks, as is the case
for diffusion in a potential landscape with several deep wells (high
barriers) \cite{eigenvalues,HTB} (see also \cite{HHS-PRE}). The
evaluation of the leading eigenvalues of the Neumann problem for the
Laplace equation in domains with bottlenecks reduces to the
computation of the leading order eigenvalue for the mixed
Neumann-Dirichlet boundary value problem for the Laplace equation in
a domain with reflecting (Neumann) boundary except for a small
absorbing (Dirichlet) window at the end of a funnel. Some estimates
on the asymptotic behavior of the leading eigenvalue are given in
\cite{Arrieta}, \cite{WardStafford} and references therein.
\subsection{Eigenvalue of the mixed problem in domains with bottlenecks}\label{ss:Bottlenecks}
\begin{figure}
\centering \resizebox{!}{6cm}{\includegraphics{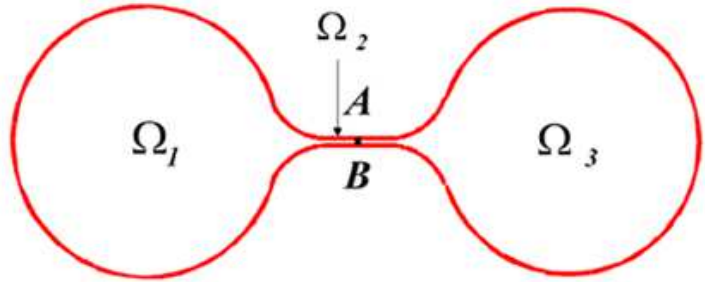}} \caption{\small A dumbbell-shaped domain
consists of two large compartments $\Omega_1$ and $\Omega_3$ connected by a narrow neck $\Omega_2$.
The bottleneck is the interval $\AB$.} \label{f:dumbbell}
\end{figure}
First we consider the principal eigenvalue of the mixed
Neumann-Dirichlet problem for the Laplace equation in a composite
domain that consists of a head $\Omega_1$ connected by a funnel to a
narrow cylindrical neck $\Omega_2$. The boundary of the domain is
reflecting (Neumann) and only the end of the cylinder $\p\Omega_a$
is absorbing (Dirichlet). The left half of Figure \ref{f:dumbbell}
shows the composite domain and the absorbing boundary is the
interval $\AB$. In the three-dimensional case the Dirichlet boundary
$\p\Omega_a$ is a small absorbing disk at the end of the cylinder.
The domain $\Omega_1$ is the one shown in Figure
\ref{f:Canonical-partial-block1} and it is connected to the cylinder
at an interface $\p\Omega_i$, which in this case is the interval
$AB$ in Figure \ref{f:Canonical-partial-block1}. It was shown in
\cite{Dire-Part-II} that the MFPT from $\x\in\Omega_1$ to
$\p\Omega_a$ is given by
\begin{align}
\ds{\bar\tau_{\small\x\to\p\Omega_a}=\bar\tau_{\small\x\to\p\Omega_i}+
\frac{L^2}{2D}+\frac{|\Omega_1|L}{|\p\Omega_a|D}}.\label{taubar}
\end{align}
The principal eigenvalue of the mixed two- and three-dimensional
Neumann-Dirichlet problems in domains with small Dirichlet and large
Neumann parts of a smooth boundary is asymptotically the reciprocal
of the MFPT given in (\ref{taubar}). Thus the principal eigenvalue
$\lambda_1$ in a domain with a single bottleneck is given by
\begin{align}
\lambda_1\sim\frac{1}{\ds{\bar\tau_{\small\x\to\p\Omega_i}+
\frac{L^2}{2D}+\frac{|\Omega_1|L}{|\p\Omega_a|D}}},\label{lambda1}
\end{align}
where $\bar\tau_{\small\x\to\p\Omega_i}$ is any one of the MFPTs given in
(\ref{taudisk})-(\ref{eq:v-sphere0}), depending on the geometry of $\Omega_1$.

If a composite domain consists of a single head and $N$ well-separated bottlenecks of different
radii and neck lengths, the derivation of (\ref{bartauN}) shows that the reciprocal of the MFPT is
the sum of the reciprocals of the NETs from a domain with a single bottleneck. That is, the
principal eigenvalue $\lambda_P$ is given by
\begin{align}
\lambda_P\sim\sum\limits_{j=1}^N\lambda_j.
\end{align}
This can be interpreted as the fact that the total efflux is the sum of $N$ independent effluxes
through the bottlenecks.
\subsection{The principal eigenvalue in dumbbell-shaped domains}
We consider now the principal eigenvalue of the Neumann problem in a
dumbbell-shaped domain that consists of two compartments $\Omega_1$
and $\Omega_3$ and a connecting neck $\Omega_2$ that is effectively
one-dimensional, such as shown in Figure \ref{f:dumbbell}, or in a
similar domain with a long neck. We assume, as we may, that the
stochastic separatrix (SS) in the neck is the cross section at its
center. In the planar case it is the segment $\AB$ in Figure
\ref{f:dumbbell}. This means that a Brownian trajectory that hits
the SS is equally likely to reach either compartment before the
other. Thus the mean time to traverse the neck from compartment
$\Omega_1$ to compartment $\Omega_3$ is asymptotically twice the
MFPT $\bar\tau_{\small\x\to SS}$ from $\x\in\Omega_1$ to the SS
\cite{equilibrium}. This MFPT is to leading order independent of
$\x\in\Omega_1$ and can be denoted $\bar\tau_{\small\Omega_1\to
SS}$.

First, we note that the mean residence time of a Brownian trajectory in
$\Omega_1$ or in $\Omega_3$ is much larger than that in $\Omega_2$ when the
neck is narrow. Second, we note that the first passage time $\tau_{\small\x\to
SS}$ for $\x\in\Omega_1$ is exponentially distributed for long times and so is
$\tau_{\small\x\to SS}$ for $\x\in\Omega_3$ \cite{DSP}. We can therefore
coarse-grain the Brownian motion to a two-state Markov process (a telegraph
process), which is in State I when the Brownian trajectory is in $\Omega_1$ and
is State II when it is in $\Omega_3$. The state $\Omega_2$ and the residence
time there can be neglected relative to those in  $\Omega_1$ and $\Omega_3$.
The transition rates from I to II and from II to I are, respectively,
\begin{align}
\lambda_{I\to II}=\frac{1}{2\bar\tau_{\small\Omega_1\to
SS}},\quad\lambda_{II\to I}=\frac{1}{2\bar\tau_{\small\Omega_3\to
SS}}.\label{rates}
\end{align}
These rates can be found from (\ref{lambda1}), with $L$ half the length of the neck and
$SS=\p\Omega_a$. The radii of curvature $R_{c,1}$ and $R_{c,3}$ at the two funnels may be
different, and the domain is either $\Omega_1$ or $\Omega_3$, as the case may be. The smallest
positive eigenvalue $\lambda$ of the Neumann problem for the Laplace equation in the dumbbell is to
leading order that of the two-state Markov process, which is $\lambda=-(\lambda_{I\to
II}+\lambda_{II\to I})$ (see Appendix below). For example, if the solid dumbbell consists of two
general heads connected smoothly to the neck by funnels (see (\ref{bartau1/2})), the two rates are
given by
\begin{align}
\frac{1}{\lambda_{I\to II}}=&\sqrt{2}\left[\left(\frac{R_{c,1}}{a}\right)^{3/2}
\frac{|\Omega_1|}{R_{c,1}D}\right](1+o(1))+\frac{L^2}{4D}+\frac{|\Omega_1|L}{\pi a^2D}\\
\frac{1}{\lambda_{II\to
I}}=&\sqrt{2}\left[\left(\frac{R_{c,3}}{a}\right)^{3/2}\frac{|\Omega_3|}{R_{c,3}D}\right](1+o(1))+\frac{L^2}{4D}
+\frac{|\Omega_3|L}{\pi a^2 D} .
\end{align}
Next, we consider the Neumann problem for the Laplace equation in a
domain that consists of any number of heads interconnected by narrow
necks. The Brownian motion can be coarse-grained into a Markovian
random walk that jumps between the connected domains at
exponentially distributed times with rates determined by the first
passage times and exit probabilities, as described in Section
\ref{ss:Bottlenecks}. This random walk can in turn be approximated
by an effective coarse-grained anisotropic diffusion, as done, for
example, for atomic migration in crystals \cite[Ch.8, Sect. 2]{book}
and for effective diffusion on a surface with obstacles
\cite{HHS-PRE}.

\section{A Brownian needle in dire straits}\label{s:Brownian-needle}
\begin{figure}[ht!]
\centering \resizebox{!}{5cm}{\includegraphics{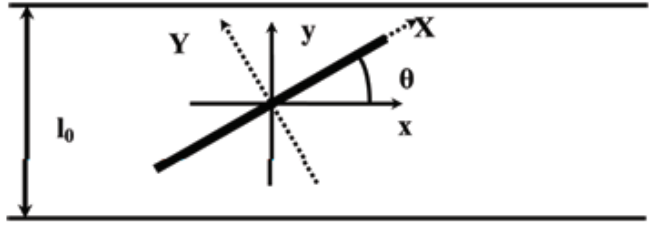}} \centering
\resizebox{!}{7cm}{\includegraphics{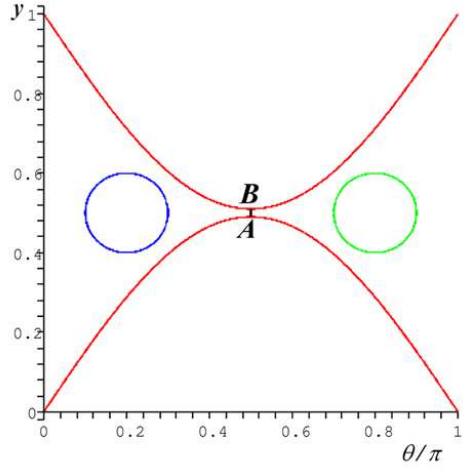}} \caption{\small {\bf Top:} Rod
in strip . The strip width is $l_0$ and the rod length is $l<l_0$. The position
of the rod is characterized by the angle $\theta$ and the fixed coordinates $x$
and $y$ and the rotating system of coordinates $(X,Y,\theta)$. {\bf Bottom:}
The motion of the rod is confined to the domain $\Omega$ in the $y,\theta$
plane.} \label{f:strip}
\end{figure}

As an application of the methodology described above, we study the planar diffusion of a stiff thin
rod (needle) of length $l$ in an infinite horizontal strip of width $l_0>l$. We assume that the rod
is a long thin right circular cylinder with radius $\epsilon\ll l_0$ (Figure \ref{f:strip}). The
planar motion of the rod is described by two coordinates of the centroid and the rotational angle
$\theta$ between the axes of the strip and the rod. The $y$-coordinate of the center of the rod is
measured from the axis of the strip. The motion of the rod is confined to the domain $\Omega$ shown
in Figure \ref{f:strip}b. The rod turns across the vertical position if it goes from the green to
the blue domains or in the reverse direction. If
 \beq
\varepsilon= \frac{l_0-l}{\l_0}\ll1,\label{l0l}
 \eeq
the black window becomes narrow and the mean first passage times (MFPT) $\tau_{\mbox{\tiny{\bf
Green$\to$Black}}}$ and $\tau_{\mbox{\tiny{\bf Blue$\to$Black}}}$, from the green or blue to the
black segment, become much longer than those in the other directions. The former also become
independent of the starting position outside a boundary layer near the black segment. Thus the
definition of the time to turn is independent of the choice of the green and blue domains, as long
as they are well separated from the black segment. The neck near the black domain is the boundary
layer region near $\theta=\pi/2$. We neglect henceforward the short times relative to the long
ones.

To turn across the vertical position the rod has to reach the black domain from the green one for
the first time and then to reach the blue domain for the first time, having returned to the green
domain any number of times prior to reaching the blue domain. Due to symmetry, a simple renewal
argument shows that the mean time to turn, $\tau_{\mbox{\tiny{\bf Blue$\to$Green}}}$, is
asymptotically given by
 \beq
 \label{2BB}
\tau_{\mbox{\tiny{\bf Blue$\to$Green}}}\sim 2\tau_{\mbox{\tiny{\bf
Blue$\to$Black}}}\quad\mbox{for}\quad\frac{l_0-l}{\l_0}\ll1.
 \eeq
The time to turn is invariant to translations along the strip (the $x$-axis), therefore it suffices
to describe the rod movement by its angle $\theta$ and the $y$ coordinate of its center. The
position of the rod is defined for $\theta\mod\pi$. Therefore the motion of the rod in the
invariant strip can be mapped into that in the $(\theta,y)$  planar domain $\Omega$ (see
Fig.\ref{f:strip}b):
 \beq
\Omega=\left\{ (\theta,y)\,:\, |y|<\frac{l_0-l\sin\theta}{2},\quad0<\theta<\pi
\right\}.\label{Omega}
 \eeq
\subsection{The diffusion law of a Brownian needle in a planar strip}
In a rotating system of coordinates $(X,Y,\theta)$, where the instantaneous $X$-axis is parallel to
the long axis of the rod and the $Y$-axis is perpendicular to it, the diffusive motion of the rod
is an anisotropic Brownian motion, and can be described by the stochastic equations
 \beqq
\dot{X}& =&\sqrt{2D_X} \dot{w}_1\\
&&\\
\dot{Y} &=&\sqrt{2D_Y} \dot{w}_2\\
&&\\
 \dot{\theta}& =&\sqrt{2D_r} \dot{w}_3,
 \eeqq
where $D_X$ is the longitudinal diffusion coefficient along the axis, $D_Y$ the transversal
diffusion constant and $D_r$, the rotational diffusion coefficient. Due to the anisotropy, the rod
makes in general larger excursions in the $X$-direction than in the $Y$-direction and this usually
characterized by the ratio $\ds\frac{D_Y}{D_X}$. In a fixed system of Cartesian coordinates
$(x,y)$, the translational and rotational motion of the centroid $(x(t),y(t))$ and the angle of
rotation $\theta(t)$ of the rod is governed by the It\^o equations
 \beqq
\dot{x}& =&\cos(\theta)\sqrt{2D_X}\, \dot{w}_1 -\sin(\theta)\sqrt{2D_Y}\, \dot{w}_2\\
&&\\
\dot{y}& =&\sin(\theta)\sqrt{2D_X}\,\dot{w}_1+\cos(\theta)\sqrt{2D_Y}\,\dot{w}_2\\
&&\\
 \dot{\theta}& =&\sqrt{2D_r} \dot{w}_3,
 \eeqq
which can be put in the matrix form
 \beqq
 \dot\x(t)=\mb{B}(\theta)\,\dot\w,
 \eeqq
where
 \beqq
\x&=&\left(\begin{array}{c}x\\y\\\theta\end{array}\right),
\quad\w=\left(\begin{array}{c}w_1\\w_2\\w_3\end{array}\right)
 \eeqq
and
 \beqq
 \mb{B}(\theta)=\sqrt{2}\left(\begin{array}{rrr}
 \cos\theta&-\sin\theta&0\\
\sin\theta& \cos\theta&0\\
0&0&1\end{array}\right)\left(\begin{array}{ccc}
 \sqrt{D_X}&0&0\\
0&\sqrt{D_Y}&0\\
0&0&\sqrt{D_r}\end{array}\right).
 \eeqq
The probability density function of the rod in the product space $\Omega \times \rR$,
 \beq
p(t,x,y,\theta)\,d\x= \Pr \{(x(t),y(t),\theta(t)) \in \x+d\x\},
 \eeq
satisfies the Fokker-Planck equation
 \beqq
\frac{\p p(t,\x)}{\p t}= -\nabla\cdot \mb{J}(t,\x),
 \eeqq
where the flux is given by
  \beq
\mb{J}(t,\x)=- \left(\begin{array}{c} \left[D_X\cos^2\theta+D_Y\sin^2\theta\right]\ds\frac{\p p}{\p
x}
 + \frac{1}{2}\left[(D_X -D_Y)\sin 2\theta\right]\ds\frac{\p p}{\p y}\\
\\
\left[D_X\sin^2\theta+D_Y\cos^2\theta\right]\ds\frac{\p p}{\p y}
 +\frac{1}{2}\left[ (D_X -D_Y)\sin 2\theta\right]\ds\frac{\p p}{\p x }\\
\\
D_r \ds\frac{\p p}{\p \theta}\end{array}\right). \label{eq:Ka1}
 \eeq
The boundary conditions are $\pi$-periodic in $\theta$, because the position of the rod
is defined modulo $\pi$ (note that $\mb{J}(t,\x)$ is $\pi$-periodic in $\theta$). This
means that the density $p(t,x,y,\theta)$ is $\pi$-periodic and the normal flux $-D_r\p
p(t,x,y,\theta)/\p \theta$ is $\pi$-antiperiodic in $\theta$.

The MFPT $\tau_{\mbox{\tiny{\bf Blue}$\to$\tiny{\bf Black}}}$ is translation-invariant with respect
to $x$ and is, therefore, the solution $u(\theta,y)$ of the boundary value problem
 \beq
D_r\frac{\p^2 u(\theta,y)}{\p \theta^2} +D_y(\theta)\frac{\p^2 u(\theta,y)}{\p
y^2}=-1\quad\mbox{for}\quad(\theta,y)\in\Omega_1,\label{pde2}
 \eeq
where $D_y(\theta)=D_X\sin^2\theta+D_Y\cos^2\theta$ and
$\Omega_1=\Omega\cap\left\{\theta<\ds\frac{\pi}{2}\right\}$, with the boundary conditions
 \beq
 \frac{\p u}{\p\tilde n}&=&0\quad\mbox{for}\quad(\theta,y)\quad\mbox{on the red
 boundary and at $\theta=0$}\label{noflux}\\
 u\left(\frac{\pi}{2},y\right)&=&0\quad\mbox{for}\quad|y|<l_0-l,\label{absorb}
 \eeq
where the co-normal derivative of $u(\theta,y)$ on the red boundary is given by
  \beq
\frac{\p u}{\p\tilde n}=\nabla
u(\theta,y)\cdot\tilde\n(\theta)\quad\mbox{for}\quad(\theta,y)\quad\mbox{on the red  boundary}
 \eeq
and the co-normal vector $\tilde\n(\theta)$ is given by
 \beq
\tilde\n(\theta)=\left(\begin{array}{cc}D_r&0\\0&D_y(\theta)\end{array}\right)\n(\theta)
 \eeq
with $\n(\theta)$ -- the unit normal vector to the red boundary.

Introducing the dimensionless variables
 \beqq
X'=\frac{X}{l_0},\quad Y'=\frac{Y}{l_0},\quad \xi(t)=\frac{x(t)}{l_0},\quad
\eta(t)=\frac{y(t)}{l_0}
 \eeqq
and the normalized diffusion coefficients
 \beqq
D'_X=\frac{D_X}{l_0^2},\quad D'_Y=\frac{D_Y}{l_0^2},\quad D_\eta(\theta)=\frac{D_y(\theta)}{l_0^2},
 \eeqq
we find that the domain $\Omega$ in (\ref{Omega}) is mapped into
 \beq
\Omega'=\left\{ (\theta,\eta) \,:\,|\eta|<\frac{1-(1-\eps)\sin\theta}{2},\quad0<\theta<\pi
\right\}.\label{Omega'}
 \eeq
To convert (\ref{pde2}) to canonical form, we introduce the variable
 \beq
 \varphi(\theta)=\int\limits_0^\theta\sqrt{\frac{D_\eta(\theta')}{D_r}}\,d\theta',\label{xi}
 \eeq
which defines the inverse function $\theta=\theta(\varphi)$, and set $u(\theta,y)=U(\varphi,\eta)$
to obtain
 \beq
U_{\varphi\varphi}(\varphi,\eta)+U_{\eta\eta}(\varphi,\eta)=U_\varphi(\varphi,\eta)\sqrt{D_r}\,
\frac{dD_\eta^{-1/2}(\theta)}{d\theta}-\frac{1}{D_\eta(\theta)}.\label{Uphiphi}
 \eeq
The domain $\Omega'$, defined in (\ref{Omega'}), is mapped into the similar domain
 \beq
 \Omega''=\left\{
(\varphi,\eta) \,:\,|\eta|<\frac{1-(1-\eps)\sin\theta(\varphi)}{2},\quad0<\varphi<\varphi(\pi)
\right\}\label{Omega''}
 \eeq
in the $(\varphi,\eta)$ plane. Because the the co-normal direction at the boundary becomes normal,
so does the co-normal derivative. It follows that the no-flux boundary condition (\ref{noflux}) and
the absorbing condition (\ref{absorb}) become
 \beq
\frac{\p U(\varphi,\eta)}{\p n}&=&0\hspace{0.5em}\mbox{for
$(\theta(\varphi),\eta)$ on $\p\Omega''$ (the red boundary in the scaled Figure \ref{f:strip}b)}\nonumber\\
\frac{\p U(0,\eta)}{\p\varphi}&=&0\hspace{0.5em}\mbox{for
$|\eta|<\ds\frac12$}\nonumber\\
U\left(\varphi\left(\frac\pi2\right),\eta\right)&=&0\hspace{0.5em}\mbox{for
$|\eta|<\ds\frac\eps2$},\label{bcphieta}
 \eeq
respectively. The gap at $\theta=\pi/2$ is preserved and the (dimensionless) radius of curvature of
the boundary at the gap is
 \beq
 R'=\frac{2D_\eta\left(\ds\frac{\pi}{2}\right)}{(1-\eps)D_r}=\frac{2D_X}{(1-\eps)l_0^2D_r}.
 \eeq
First, we simplify (\ref{Uphiphi}) by setting
 \beq
g(\varphi)=\sqrt{D_r}\, \frac{dD_\eta^{-1/2}(\theta)}{d\theta},\quad
U(\varphi,\eta)=f(\varphi)V(\varphi,\eta)\label{gf}
 \eeq
and choosing $f(\varphi)$ such that $f'(\varphi)=\frac12f(\varphi)g(\varphi)$. Note that
 \beq
\left.\frac{dD_\eta^{-1/2}(\theta)}{d\theta}\right|_{\theta=0,\pi/2,\pi}=0.\label{dDdth}
 \eeq
Equation (\ref{Uphiphi}) becomes
 \beq
V_{\varphi\varphi}+V_{\eta\eta}=\frac{1}{f(\varphi)}\left\{\left[g(\varphi)f'(\varphi)-f''(\varphi)\right]V-
\frac{1}{D_\eta(\theta(\varphi))}\right\}.\label{Vphiphi}
 \eeq
Next, we move the origin to the center of curvature of the lower boundary by setting
$$\zeta=-\left(\eta-R'-\frac\eps2\right)+i\left[\varphi-\varphi\left(\frac\pi2\right)\right]$$  and
use the conformal mapping (\ref{w}),
\begin{align}
\omega=\frac{\zeta-R'\alpha}{R'-\alpha \zeta},\label{omega}
\end{align}
with $\omega=\rho e^{i\psi}$. We also have
\begin{align}
w'(\zeta)=&\,\frac1{R'}\,\frac{(1+\alpha w)^2}{1-\alpha^2}\label{w'w}\\
|w'(\zeta)|^2=&\,\frac1{R'^2}\,\left|\frac{(1+w\alpha)^2}{1-\alpha^2}\right|^2=\frac{|1-w+\sqrt{\eps}\,w|^4}{4\eps
R'^2} (1+O(\sqrt{\eps}))\label{w'2wp},
 \end{align}
\begin{figure}[ht!]
\centering \resizebox{!}{5cm}{\includegraphics{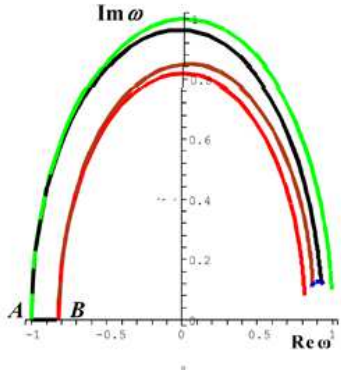}} \caption{\small The
image $\Omega_\omega$ of the domain $\Omega$ under the mapping (\ref{omega}).
The values of the parameters are $\eps=0.01$ with the approximation $D_Y\ll
D_X$. The domain is enclosed by the real segment AB and by the brown, black,
and blue curves. The green and red curves are the images of arcs of the
osculating circles at the narrow neck, as in Figure \ref{f:Rod-domain}.}
\label{f:sin-map}
\end{figure}
The image $\Omega_\omega$ of the domain $\Omega$ is given in Figure \ref{f:sin-map} and is similar
to  $\Omega_w$ in Figure \ref{f:Rod-domain}, except for a small distortion near
$\psi=c\sqrt{\eps}$, which we neglect, as we may. Setting $V(\varphi,\eta)=W(\rho,\psi)$, fixing
$\rho=1$ in $\Omega_\omega$, as in Section \ref{s:Dire straits}, and abbreviating $W=W(\psi,1)$,
equation (\ref{Vphiphi}) becomes to leading order
 \beq
 W_{\psi\psi}+\frac{h(\psi)}{|\omega'(\zeta|^2}W=-\frac{1}{|\omega'(\zeta)|^2k(\psi)},\label{Wpsipsi}
 \eeq
where
 \beq
 h(\psi)=\left.\frac{f''(\varphi)-g(\varphi)f'(\varphi)}{f(\varphi)}\right|_{\rho=1},\quad
 k(\psi)=f(\varphi)D_\eta(\theta(\varphi))|_{\rho=1}.
 \eeq
Using (\ref{PDEw}) and neglecting terms of order $O(\eps)$, we rewrite (\ref{Wpsipsi}) as
 \beq
 W_{\psi\psi}+\frac{4\eps R'^2h(\psi)}{| e^{i\psi}(1-\sqrt{\eps})-1|^4}W=
 -\frac{4\eps R'^2}{| e^{i\psi}(1-\sqrt{\eps})-1|^4k(\psi)}.\label{WpsipsiO}
 \eeq
In view of (\ref{dDdth}), the boundary conditions (\ref{bcphieta}) become
 \beq
 W_\psi(c\sqrt{\eps})=0,\quad W(\pi)=0.\label{bcW}
 \eeq
\subsection{The asymptotic solution}
The construction of the asymptotic expansion of the solution of the boundary
layer equation (\ref{WpsipsiO}) is similar to that in Section
\ref{ss:Asymptotic}. The outer solution of (\ref{WpsipsiO}) is a linear
function $W_{\mbox{\scriptsize outer}}(\psi)=a\psi+b$, where $a$ and $b$ are
yet undetermined constants. The uniform approximation is constructed as
$W_{\mbox{\scriptsize uniform}}(\psi)=W_{\mbox{\scriptsize
outer}}(\psi)+W_{\mbox{\scriptsize bl}}(\psi)$, where the boundary layer
$W_{\mbox{\scriptsize bl}}(\psi)$ is a function $Y(\xi)$ of the boundary layer
variable $\xi=\psi/\sqrt{\eps}$. The boundary layer equation is
 \beq
 Y''(\xi)+\frac{4R'^2h(0)}{(1+\xi^2)^2} Y(\xi)=-\frac{4R'^2}{(1+\xi^2)^2k(0)},
 \eeq
which is simplified by the substitution $Y(\xi)=\tilde Y(\xi)+1/h(0)k(0)$ to
 \beq
\tilde Y''(\xi)+\frac{4R'^2h(0)}{(1+\xi^2)^2} \tilde
Y(\xi)=0.\label{BLEQ2}
 \eeq
The boundary conditions (\ref{bcW}) become $\tilde Y'(c)=0$ and
$\tilde Y(\infty)=1/h(0)k(0)$. The boundary layer equation
(\ref{BLEQ2}) has two linearly independent solutions, $\tilde
Y_1(\xi)$ and $\tilde Y_2(\xi)$, which are linear for sufficiently
large $\xi$. Initial conditions for $\tilde Y_1(\xi)$ and $\tilde
Y_2(\xi)$ can be chosen so that $\tilde Y_2(\xi)\to const$ as
$\xi\to\infty$ (e.g., $\tilde Y_2(0)=-4.7,\ \tilde Y_2'(0)=-1$, see
Figure \ref{f:Boundary-layers}). Thus the boundary layer function is
given by
\begin{align}
W_{\mbox{\scriptsize bl}}(\psi)=A\tilde Y_1\left(\frac{\psi}{\sqrt{\eps}}\right) +B\tilde
Y_2\left(\frac{\psi}{\sqrt{\eps}}\right)+C,
\end{align}
where $A$ and $B$ are constants to be determined and $C$ is related to the constant $1/h(0)k(0)$
and is also determined below from the boundary and matching conditions.

The matching condition is that $W_{\mbox{\scriptsize bl}}(\psi)=A\tilde
Y_1\left(\psi/\sqrt{\eps}\right)+ B\tilde Y_2\left(\psi/\sqrt{\eps}\right)+C$ remains bounded as
$\xi\to\infty$, which implies $A=0$. It follows that at the absorbing boundary $\psi=\pi$ we have
\begin{align}
W_{\mbox{\scriptsize unif}}(\pi)=&\,a\pi+b'=0\label{absBC2}
\\
W_{\mbox{\scriptsize unif}}'(\pi)=&\,a.\nonumber
\end{align}
where the constant $b'$ incorporates all remaining constants. At the reflecting boundary we have to
leading order
\begin{align}
W_{\mbox{\scriptsize unif}}'(c\sqrt{\eps})=&\,W_{\mbox{\scriptsize
outer}}'(c\sqrt{\eps})+W_{\mbox{\scriptsize bl}}'(c\sqrt{\eps}) =
a+B\frac{\tilde Y_2'(c)}{\sqrt{\eps}}=0,\label{ypu2}
\end{align}
which gives
\begin{align}
B=-\frac{a\sqrt{\eps}}{\tilde Y_2'(c)},\quad b'=-a\pi.
\end{align}
The uniform approximation to $W(\omega)$ is given by
\begin{align}
W_{\mbox{\scriptsize unif}}(\rho e^{i\psi})=a\left(\psi-\pi-\frac{\sqrt{\eps}}{\tilde
Y_2'(c)}\right),\label{vunif}
\end{align}
so that using (\ref{gf}), (\ref{dDdth}), and (\ref{w'w}), we obtain from (\ref{vunif})
\begin{align}
\left.\frac{\p u}{\p
n}\right|_{\zeta\in\p\Omega_a}=f\left(\varphi\left(\frac{\pi}{2}\right)\right)\left.\frac{\p W(\rho
e^{i\psi})}{\p
\psi}\right|_{\psi=\pi}\omega'(\zeta)\Big|_{\zeta=-1}\left.\frac{\p\varphi}{\p\theta}\right|_{\theta=\pi/2}=
a\sqrt{\frac{2}{\eps R'}}(1+O(\sqrt{\eps})).\label{dudn}
\end{align}
Because $W(\omega)$ scales with $1/f(\varphi)$ relative to $V(\varphi,\eta)$, we may choose at the
outset $f(\varphi(\pi/2))=1$.

Finally, to determine the value of $a$, we integrate (\ref{pde2}) over
$\Omega$, use (\ref{dudn}), and the fact that
$$\int\limits_{\p\Omega_a}\,dy=l_0\eps,$$ to obtain
$a=-|\Omega|\sqrt{R'}/l_0D_r\sqrt{2\eps}$. Now (\ref{vunif}) gives the MFPT at any point $\x$ in
the head as
\begin{align}
E[\tau\,|\,\x]=u(\x)\sim W\left(\rho
e^{ic\sqrt{\eps}}\right)\sim-a\pi=\frac{\pi|\Omega|\sqrt{R'}}{l_0D_r\sqrt{2\eps}}(1+O(\sqrt{\eps})\hspace{0.5em}\mbox{for}
\ \eps\ll1.
\end{align}
Reverting to the original dimensional variables, we get
\begin{align}
E[\tau\,|\,\x]=\frac{\pi\left(\ds\frac{\pi}{2}-1\right)}{D_r\sqrt{l_0(l_0-l)}}
\sqrt{\frac{D_X}{D_r}}\left(1+O\left(\sqrt{\frac{l_0-l}{l_0}}\right)\right),
\end{align}
which is (\ref{turnaround}).
\section{Discussion and conclusion}
This paper develops a boundary layer theory for the solution of the mixed
Neumann-Dirichlet problem for the Poisson equation in geometries in which the
methodologies of \cite{Ward1}-\cite{PNAS} fail. These methodologies were used
for the narrow escape problem. In the geometries considered here the small
Dirichlet part is located at the end of narrow straits connected smoothly to
the Neumann boundary of the domain. Additional problems related to Brownian
motion in composite domains that contain a cylindrical narrow neck connected
smoothly or sharply to the head are considered in \cite{Dire-Part-II}. These
include the asymptotic evaluation of the NET, of the leading eigenvalue in
dumbbell-shaped domains and domains with many heads interconnected by narrow
necks, the escape probability through any one of several narrow necks, and
more.

Our results have applications in several areas. The first application is in
neuroscience and concerns dendritic spines, which are believed to be the locus
of postsynaptic transmission. Recognized more than 100 years ago by Ram\'on y
Cajal, dendritic spines are small terminal protrusions on neuronal dendrites,
and are the postsynaptic parts of excitatory synaptic connections. The spine
consists of a relatively narrow cylindrical neck connected to a bulky head. The
geometrical shape of a spine correlates with its physiological function
\cite{Harris0}-\cite{Newpher}. Several physiological phenomena are regulated by
diffusion in dendritic spines. For example, synaptic plasticity is induced by
the transient increase of calcium concentration in the spine, which is
regulated by spine geometry, by endogenous buffers, and by the number and rates
of exchangers \cite{EJN,Yuste,DenkSvonoda1996,PRE2007Armin,KupkaHolcman2010}.
Another significant function of the spine is the regulation of the number and
type of receptors that contribute to the shaping of the synaptic current
\cite{Choquet2}-\cite{CEllCAlcium}. Indeed, the neurotransmitter receptors,
such as AMPA and NMDA, whose motion on the spine surface is diffusion, mediate
the glutamatergic-induced synaptic current. Thus dendritic spines regulate both
two-dimensional motion of neurotransmitter receptors on its surface, and the
three-dimensional diffusive motion of ions (e.g., calcium), molecules, proteins
(e.g., mRNA), or small vesicles in the bulk. Our results give a quantitative
measure of the effect of geometry on regulation of flux.

In a biochemical context, the NET (eq. \ref{bartau1/2f}) accounts
for the local geometry near an active binding site occluded by the
molecular structure of the protein. This is the case of proton
binding sites located on spike proteins, located on the viral
envelope  of the influenza virus and involved in membrane fusion
\cite{Herrmann}. Another application is that of the turnaround time
of Brownian needle. Our result in Section \ref{s:Brownian-needle}
provides the precise time scale of the unraveling of a double strand
DNA break confined between two-dimensional membranes \cite{Minsky}.
The common feature of the geometries studied in this paper is the
cusp-shaped narrow passage leading to the absorbing boundary. The
main biological conclusion of our results is that this geometry is
the main controller of the flux through biological narrow passages,
an effect that is ubiquitous in biological systems. More specific
applications of the NET from composite domains to dendritic spines
are given in \cite{Dire-Part-II}.

\section{Appendix}
The asymmetric random telegraph process jumps between two states, $a$ and $b$,
at independent exponentially distributed waiting times with rates
$\lambda_{a\to b}$ and $\lambda_{b\to a}$, respectively. The transition
probability distribution function satisfies the linear differential equations
(see http://en.wikipedia.org/wiki/Telegraph$\_{}$process, \cite{DSP})
\begin{align}
\frac{\p P\{a,t\,|\,x,t_0\}}{\p t}=&\,-\lambda_{a\to
b}P\{a,t\,|\,x,t_0\}+\lambda_{b\to a}P\{b,t\,|\,x,t_0\}\nonumber\\
&\label{system}\\
 \frac{\p P\{b,t\,|\,x,t_0\}}{\p t}=&\,\lambda_{a\to
b}P\{a,t\,|\,x,t_0\}-\lambda_{b\to a}P\{b,t\,|\,x,t_0\},\nonumber
\end{align}
which can be written in the obvious matrix notation as $\dot{\mb{p}}=\A\mb{p}$
with
\begin{align*}
\A=\left(\begin{array}{rr}-\lambda_{a\to b}&\lambda_{b\to a}\\ \lambda_{a\to
b}&-\lambda_{b\to a}\end{array}\right).
\end{align*}
The eigenvalues of $\A$ are $0$ with the normalized eigenvector
$(\frac12,\frac12)^T$, and $-(\lambda_{a\to b}+\lambda_{b\to a})$ with the
eigenvector $(1,-1)^T$. It follows that the nonzero eigenvalue of the system
(\ref{system}) is $\lambda=\lambda_{a\to b}+\lambda_{b\to a}$.\\

\noindent {\bf Acknowledgment:} The authors wish to thank F. Marchesoni for
pointing out the factor 1/2 in eq.(\ref{bartaud0}).


\end{document}